\documentclass[11pt,reqno,letterpaper]{article}
\usepackage{geometry}
\usepackage{array}
\usepackage{xspace}

\usepackage{bbold}
\newcommand{\bbbone}{\ensuremath{\mathbb 1}}

\usepackage{amsmath,amssymb,amsthm}
\usepackage{enumerate}
\theoremstyle{plain}

\newtheorem{lemma}{Lemma}
\newtheorem{theorem}[lemma]{Theorem}
\newtheorem{corollary}[lemma]{Corollary}
\newtheorem{proposition}[lemma]{Proposition}

\theoremstyle{definition}
\newtheorem{definition}[lemma]{Definition}
\newtheorem{remark}[lemma]{Remark}

\newcounter{claim}[lemma]

\newenvironment{claim}{

\vspace{1mm}
\refstepcounter{claim}\noindent\emph{Claim \theclaim.} }{\vspace{1mm}

}
\newcounter{invariant}[lemma]

\newenvironment{invariant}{

\vspace{1mm}
\refstepcounter{invariant}\emph{Invariant \theinvariant.} }{
\vspace{1mm}

}

\newcommand{\smallqed}{\hfill\scriptsize\ensuremath{\blacksquare}

}
\newenvironment{pf}{\noindent\emph{Proof:} }{

}

\newenvironment{procedure}[2]{%
\vspace*{1mm}\\
\noindent\mbox{\textbf{Procedure} \textsc{#1} (#2)}%
\begin{enumerate}[(1.)]%
\itemsep=0pt\topsep=0pt%
\partopsep=0pt\raggedright\parskip=2pt\parsep=0pt%
\vspace{-2mm}
}{\end{enumerate}}

\newcommand{\Vout}{V_{out}}
\newcommand{\tent}{\mathit{tent}}
\newcommand{\Medium}{\mathit{Medium}}
\newcommand{\Pend}{\mathit{Pending}}
\newcommand{\F}{\mathcal{F}}
\newcommand{\Finish}{\mathit{Finished}}
\newcommand{\Heavy}{\mathit{Heavy}}
\newcommand{\Light}{\mathit{Light}}
\newcommand{\Untouch}{\mathit{Untouched}}
 
\newcommand{\ichi}{\chi^{\text{-}1}}
\newcommand{\B}{\mathcal B}
\newcommand{\Vin}{V_{in}}
\newcommand{\bnd}{\partial}
\newcommand{\Shrink}{\textsc{Shrink}\xspace}
\newcommand{\Move}{\textsc{Move}\xspace}
\newcommand{\floor}[1]{\lfloor #1 \rfloor}
\newcommand{\exs}{\mathit{excess}}
\newcommand{\Add}{\textsc{AddTo}\xspace}
\newcommand{\Del}{\textsc{CutDown}\xspace}
\newcommand{\Empty}{\textsc{ReduceBuffer}\xspace}
\newcommand{\hichi}{\hat\chi^{\text{-}1}}

\newcommand{\lfluc}{\phi_\ell}
\newcommand{\tichi}{\tilde\chi^{\text{-}1}}
\newcommand{\Buffer}{\mathit{Buffer}}
\newcommand{\eq}[1]{\begin{equation}#1\end{equation}}
\newcommand{\tV}{ V_{in}}

\newcommand{\col}{\colon}
\newcommand{\rem}[1]{\hfill\emph{// #1}}
\newcommand{\irem}[1]{\item[\emph{//}]\emph{#1}}
\newcommand{\enum}[1]{\begin{enumerate}#1\end{enumerate}}
\newcommand{\ra}{\rightarrow}
\newcommand{\Rnn}{\mathbb R_{+}}	
\newcommand{\Rp}{\mathbb R_{>0}}	
\newcommand{\R}{\mathbb R}	
\newcommand{\N}{\mathbb N}
\newcommand{\Z}{\mathbb Z}
\newcommand{\la}{\leftarrow}
\renewcommand{\vec}[1]{\mathbf{#1}}

\newcommand{\norm}[1]{\| #1 \|}
\newcommand{\ho}[1]{^{(#1)}}
\newcommand{\infnorm}[1]{\|#1\|_\infty}
\newcommand{\onorm}[1]{\|#1\|_1}
\newcommand{\pnorm}[1]{\|#1\|_p}
\newcommand{\res}[1]{_{|#1}}
\newcommand{\avg}[1]{\norm{#1}_{avg}}
\newcommand{\Sink}{\mathit{Sink}}
\newcommand{\Source}{\mathit{Source}}
\renewcommand{\inf}[1]{\infnorm{#1}}
\newcommand{\hide}[1]{}

\title{%
  Tight Bounds on the Min-Max Boundary\\ 
  Decomposition Cost of Weighted Graphs\\
}

\date{\small\today}

\author{%
David Steurer\\
\small {MPI Informatik}\\
\small {Stuhlsatzenhausweg 85}\\
\small {66123 Saarbr\"{u}cken, Germany}\\
\small \texttt{dsteurer@mpi-inf.mpg.de}
}

\begin{document}

\maketitle

\vspace{-10mm}

\begin{abstract}
  \small Many load balancing problems that arise in scientific
  computing applications boil down to the problem of partitioning a
  graph with weights on the vertices %
  and costs on the edges into a given number of equally-weighted parts
  such that the \emph{maximum boundary cost} over all parts is small.

  Here, this partitioning problem is considered for %
graphs~$G=(V,E)$ with edge costs $c\col E\ra \Rnn$,
that have bounded maximum degree and a \mbox{\emph{$p$-separator theorem}} for some $p>1$, 
i.e., any (arbitrarily weighted) subgraph of $G$ can be separated into 
two parts of roughly the same weight
by removing a \emph{separator} $S\subseteq V$
such that the edges incident to $S$ in the subgraph
have total cost at most proportional to $(\sum_e c^p_e)^{1/p}$, 
where the sum is over all edges in the subgraph.
  
For arbitrary weights $w\col V\ra \Rnn$, we show that the vertices of
such graphs
can be partitioned into $k$ parts such that the weight of each part
differs from the average weight $\sum_{v\in V}w_v/k$ by at most
$(1-\frac{1}{k})\max_{v\in V}w_v$, and the boundary edges of each part
have total cost at most proportional to $(\sum_{e\in
  E}c_e^p/k)^{1/p}+\max_{e\in E}c_e$.  The partition can be computed
in time nearly proportional to the time for computing separators $S$
for $G$ as above.

Our upper bound is shown to be tight up to a constant factor for
infinitely many instances
with a broad range of parameters.  Previous results achieved this
bound only if one has $c\equiv 1$, $w\equiv 1$, and one allows parts
of weight as large as a constant multiple of the average weight.

We also give a separator theorem for $d$-dimensional grid graphs with
arbitrary edge costs, which is the first result of its kind for
non-planar graphs.
\end{abstract}

\section{Introduction}
\label{sec:introduction}

We consider the problem to partition a weighted graph into a given
number of parts subject to the constraint that the weight of each part
differs from the average part weight only by a relatively small
quantity.  The objective is to minimize the \emph{maximum boundary
  cost} over all parts, where the boundary cost of a part is the the
total cost of the edges with exactly one endpoint in the part.

This problem naturally arises as a load balancing problem in
scientific computing applications, where one wants to solve a
large-scale problem given by a set $V$ of jobs on a parallel computing
system with $k$ identical machines.  The weight~$w_u$ is proportional
to the time a machine takes to process job~$u\in V$.  However, job $u$
may depend on other jobs $v\in V$.  For each such dependency the
graph~$G=(V,E)$ contains an edge $e=\{u,v\}\in E$.  If job~$v$ is not
scheduled on the same machine as job $u$ then a cost~$c_e$ is induced
on the machines that handle jobs $u$ and $v$.  The cost $c_e$ reflects
the overhead for the communication needed to resolve the dependency
among jobs $u$ and $v$.  How the makespan of a schedule increases
under large communication costs, depends on the specific design of the
considered parallel computing system.  In general, one requires from a
good schedule that the weights of the jobs are as equally distributed
among the machines as possible and that the maximum communication cost
over all machines is small.  So this load balancing problem
corresponds to the graph partitioning problem from above.
For example, consider the problem of large-scale climate simulation,
where the surface of the earth is subdivided into many triangular
regions.  For each region, there is a job in $V$ to simulate the
weather in this region for a period of time.  Of course, the
simulations for neighboring regions depend on each other.  So if jobs
of neighboring regions are scheduled on distinct machines, one might
have to interchange considerable amounts of data between the machines
causing an increase of the makespan.  This example also illustrates
the use of weights and costs.  Even if all regions have about the same
area, the time for simulating the weather in these regions might
differ tremendously depending on day-time, desired accuracy, et
cetera.  The degree of dependency among neighboring regions might
differ in a similar manner.

Our aim is to characterize graph classes that, even for worst possible
weights, allow $k$-way partitions that are good in the sense above,
i.e., have equally-weighted parts and small boundary costs.  We shall
see that a ``well-behaved'' graph class allows good $k$-way partitions
if and only if it allows good $2$-way partitions, i.e., it has a
\emph{separator theorem}.  In this, we can predict the scalability of
the mentioned scientific computing applications.

Our results imply that there is no inherent trade-off between the
weight-balanced\-ness of a partition and its boundary costs.  In
particular, any partition, with weight of each part at most
proportional to the average, can be transformed into a partition with
almost equally-weighted parts such that the maximum boundary cost
increases by at most constant factor, essentially.

Notice that the ``quality'' of a partition could also be measured by
the average boundary cost instead of the maximum boundary cost.  One
might ask whether there are considerably better upper bounds %
for this measure than for the maximum boundary cost.  We answer this
question in the negative.
\subsubsection*{Previous Work and Contributions}
Much work has been done on worst-case guarantees for graph
partitioning problems.  In a seminal article, Lipton and Tarjan
\cite{tarjan:sep} established a separator theorem for planar graphs,
asserting that every $n$-vertex planar graphs can be separated into
two parts of size at most $2n/3$ by removing $O(n^{1/2})$ vertices.
Further separator theorems exist for graphs with an excluded minor
\cite{alon:sep} and for $d$-dimensional well-shaped meshes
\cite{teng:mesh,spielman:mesh,miller:sep}, where for the latter
$O(n^{1-1/d})$ vertices can be removed instead of $O(n^{1/2})$.  More
generally, a graph is said to have a \emph{$p$-separator theorem}
(with respect to unit costs) if any induced subgraph can be separated
into two parts of about the same weight by removing $O(n_0^{1/p})$
vertices, where $n_0$ is the number of vertices in the subgraph.

Simon and Teng \cite{simon97how} addressed the problem to partition a
graph into $k\ge 2$ parts of weight at most proportional to the
average, by removing edges from the graph.  They showed that for
bounded-degree graphs with a $p$-separator theorem such a partition
can be achieved by removing $O(k^{1-1/p} n^{1/p})$ edges.  So for unit
edge-costs,%
the average boundary of the partition is at most proportional to
$(n/k)^{1/p}$.

Kiwi, Spielman and Teng \cite{spielman:minmax} were the first to give
bounds on the maximum boundary cost instead of the average boundary
cost.  For unit-weights and unit-costs, they show that bounded-degree
graphs with $n$ vertices and $p$-separator theorem can be decomposed
into $k$ parts such that the weight of each parts is $O(n/k)$ and the
maximum boundary cost is at most proportional to $(n/k)^{1/p}$.  They
also give bounds for partitions with maximum weight at most
$(1+\epsilon)\cdot n/k$ and for the case of arbitrary weights.
However, in these cases their bound on the maximum boundary cost
increases by a factor $(1/\epsilon)^{1-1/p}$ and
$(\log(k/\epsilon^{2})/\epsilon)^{2-2/p}$, respectively.  We show that
this asymptotic increase of the maximum boundary cost can be avoided.
More specifically, our bounds for the weighted case are the same as
for the unweighted case, and in our results there is no trade-off
between balancedness and boundary costs.

In Appendix~\ref{lowersec} we show that the obtained worst-case bounds
on the maximum boundary cost are optimal with respect to the chosen
parameters.

\paragraph*{Strict weight-balancedness.}
It seems new to allow the constraint that the weight of each part may
differ from the average weight of a part by at most
$\frac{k-1}{k}\max_{v\in V} w_v$.  Notice that this guarantee on the
weight of the parts is the same as of an algorithm that assigns each
vertex greedily to a part, i.e., a greedy bin-packing algorithm.
However,
in contrast to our methods, such a greedy algorithm will in general
create huge boundary costs.  In Section~\ref{sec:shrink} we present a
novel ``shrink-and-conquer'' algorithm that transforms any partition
with loosely balanced weights into a strictly weight-balanced
partition while maintaining the bounds on the maximum boundary
cost. For the conquer-phase, a greedy bin-packing procedure is used.
But our ``shrink-and-conquer'' approach shall ensure that this packing
procedure touches every part only constantly often and therefore the
boundary costs do increase only slightly in a conquer-phase.

\paragraph*{Arbitrary edge costs.} 
If one allows arbitrary costs $c\col E\ra \Rnn$ on the edges instead
of unit costs, then only the separator theorem for planar graphs
\cite{djidjev:cost} was known to extend to this case. Any
bounded-degree planar graph can be separated into two parts of about
the same weight by removing edges of cost $O((\sum c_e^2)^{1/2})$.  In
Section~\ref{sec:splitt-grid-graphs}
we give a separator theorem for $d$-dimensional grid graphs.  Every
$d$-dimensional grid graph can be separated into two almost
equally-weighted parts by removing edges of cost at most proportional
to $(\sum c_e^{d/(d-1)})^{1-1/d}\cdot \log^{1/d}\phi$, where
$\phi:=\max_e c_e/\min_e c_e$ is the fluctuation of the edge costs.
We think that the logarithmic factor in our grid separator theorem is
superfluous.  Moreover, we conjecture that many graph classes with
separator theorem for unit costs also have a separator theorem for
arbitrary $c$.

Assuming such separator theorems, we can extend the bounds on the
maximum boundary cost of $k$-way partitions to the case of arbitrary
edge costs.  For this generalization, we utilize multi-balanced
partitions (cf. Section~\ref{multbalsec}), i.e., partitions that are
simultaneously balanced with respect to several weight functions.
Multi-balanced partitions were implicitly considered by Kiwi,
Spielman, and Teng \cite{spielman:minmax}. Their idea is to use
recursive bisection where each separator divides the vertices evenly
with respect to all weight functions.
Such separators are increasingly difficult to find when the number of
weight functions grows larger.
Their approach gives the same guarantee as ours only if there are at
most two weight functions.
We use, instead of ordinary recursive bisection, a generalization
thereof, which allows to balance the partition with respect to the
weight functions one by one.
This approach makes it possible to handle an arbitrary number of
weight functions.

\subsubsection*{Notation}
In this work, all considered graphs are assumed to be finite,
undirected, and without self-loops or parallel edges.  A graph
$G=(V,E)$ has \emph{size} $|G|:=|V|+|E|$.  For a subset $U\subseteq V$
of the vertices, $\delta(U):=\{e\in E\mid |e\cap U|=1\}$ denotes the
\emph{cut} induced by $U$, or the set of \emph{boundary edges} of $U$.
We let $G[W]:=(W,E(W))$ be the graph induced by a vertex set $W
\subseteq V$ in $G$, where $E(W):=\{e\in E\mid e\subseteq W\}$ is the
set of edges running in $W$.  For all other graph notations, we refer
to any standard text book on graph theory or algorithms.

Let $f\col X\ra \Rnn$ be a function on a finite domain $X$.  If not
ambiguous, we write $f_x:=f(x)$.  For a subset $S\subseteq X$, we
define $f(S):=\sum_{s\in S}f_s$.  For $p>1$, the \emph{$p$-norm} of
$f$ is given by $\pnorm f:=(\sum_{x\in X}f^p_x)^{1/p}$.  In the limit,
we have $\inf f=\max_{x\in X}f_x$.
\emph{H\"older's inequality} states that $\sum_{x\in X}f_xg_x\le
\pnorm f\cdot \norm g_q$ for functions $f,g\colon X\ra \Rnn$ and
$p,q>1$ with $\frac{1}{p}+\frac{1}{q}=1$.  We denote the
\emph{restriction} of $f$ to $S\subseteq X$ by $f\res{S}\colon S\ra
\Rnn$ with $f\res S(x):=f(x)$ for all $x\in S$. The function on domain
$X$ that is identical to $1$ is denoted by $\bbbone_X:X\ra
\{1\}$. When adding two non-negative function $f$ and $g$ with
respective domains $X$ and $Y$, we implicitly extend $f$ and $g$ to
the domain $X\cup Y$ with $f\res{Y\backslash X}\equiv 0$ and
$g\res{X\backslash Y}\equiv 0$. Then we can define $f+g:X\cup Y\ra
\Rnn$ by $(f+g)(x)=f(x)+g(x)$.  When the domains $X$ and $Y$ are
disjoint we call the sum $f+g$ \emph{direct} and write sometimes
$f\oplus g$.

We write $f=O_d(g)$, $f\ll_d g$, or $g=\Omega_d(f)$ for expressions
$f$ and $g$ if $|f|\le C\cdot |g|$ for some constant $C$ that might
depend on a parameter $d$.

\section{Min-Max Boundary Decomposition Problem}
In the following we provide a formalization of this problem and state
our main results.
The decomposition problem is formulated in terms of vertex colorings
instead of partitions, since we find this notation more convenient.

\begin{definition}[Strictly Balanced Colorings] \label{def:strictcol}
  Let $k$ be a positive integer and $G=(V,E)$ be a graph with edge
  costs $c\colon E\ra \Rnn$ and vertex weights $w\colon V\ra \Rnn$.
  
  A \emph{$k$-coloring} $\chi\col V\ra [k]$ of $G$ is \emph{strictly
    ($w$-)balanced} if the weight of each \emph{color class}
  $\ichi(i):=\{v\in V\mid \chi(v)=i\}$ differs from the average weight
  of a color class by no more than a $(1-\frac{1}{k})$-fraction of
  $\infnorm w$, i.e., when we have
  \begin{equation}
    \max_{1\le i\le k}
    \left|w(\ichi(i))-\frac{\onorm w}{k}\right|
    \le (1-1/k) \cdot\infnorm w.\label{eq:0}
  \end{equation}
  The \emph{maximum boundary cost} of a coloring $\chi$ of $G$ is
  defined as the maximum cost of the boundary edges $\delta(\ichi(i))$
  of a color class $\ichi(i)$, formally,
  \[
  \infnorm{\bnd\ichi}:=\max_{1\le i\le k}c(\delta(\ichi(i))).
  \]
  (The strange symbol $\inf{\bnd\ichi}$ will be consistent with our
  further notation.)
\end{definition}

Notice that strictly balanced coloring are as weight-balanced as
possible for many parameter choices.  More precisely, for all choices
of $k$ and $\inf w$, there are infinitely many choices of $\onorm w$
such that equality holds in \eqref{eq:0} even for the most
weight-balanced coloring of some instance with the chosen parameters.

For the applications mentioned in Section~\ref{sec:introduction}, it
is desirable to know which graphs allow strictly balanced
$k$-colorings of small maximum boundary cost even if the weights of
the vertices are chosen adversarial.
\begin{definition}\label{def:decompcost}
  The \emph{min-max boundary ($k$-)decomposition cost} of $G$ with
  edge costs $c\colon E\ra \Rnn$ is the minimum maximum boundary cost
  over all strictly balanced $k$-colorings of $G$ with respect to
  worst possible weights, formally,
  \[
  \bnd^k_\infty(G,c):=\sup_{w\colon V\ra \Rnn}\min_\chi
  \infnorm{\bnd\ichi}, \] where the minimum is over all strictly
  $w$-balanced $k$-colorings $\chi$ of $G$.
\end{definition}
In our main theorem we will upper bound $\bnd^k_\infty(G,c)$ in terms
of a parameter related to~$\bnd^2_\infty$ that is subgraph-monotone,
i.e., it does not increase when going to (induced) subgraphs.  Note
that the trivial subgraph-monotone version of $\bnd^2_\infty$, namely
\[
\max_{W\subseteq V}\bnd^2_\infty(G[W],c_{|E(W)}),
\] is pointless, since it gives no information about how large the
costs for decomposing $G[U]$ are compared to the edge costs
$c_{|E(U)}$ in this subgraph.  For a meaningful parameter, we need to
relate the decomposition cost of a subgraph to the costs of its edges.

The following definition formalizes this idea.
\begin{definition}[Splitting Sets, $p$-Splittablity]
  \label{def:split}
  For any \emph{splitting value} $w^*$ with $0\le w^*\le \onorm w$, a
  vertex set $U\subseteq V$ is said to be \emph{$w^*$-splitting} if $
  |w(U)-w^*|\le \inf w/2$.  The \emph{(boundary) cost} of $U$ is $
  \bnd U:=c(\delta(U))$.

  Now the \emph{$p$-splittability} of a graph $G$ with edge costs $c$
  is the least number $\sigma_p(G,c)$ such that for every induced
  subgraph $G[W]$, all weights $w\col W\ra \Rnn$ and splitting values
  $w^*$, there exists a $w^*$-splitting set $U\subseteq W$ with
  boundary cost $\bnd_W U\le \sigma_p(G,c)\cdot \pnorm{c\res{W}}$,
  where $\bnd_W U$ is the boundary cost of $U$ in $G[W]$, and
  $c\res{W}$ denotes the restriction of $c$ to the edges of $G$
  running in $W$.
  We write $\sigma_p:=\sigma_p(G,c)$ when $G$ and $c$ are understood.
\end{definition}

We remark that if instance $(G,c)$ is \emph{well-behaved}, i.e., $G$
has bounded maximum degree and $c(u,v)=\Omega(c(u,v'))$ for all edges
$\{u,v\},\{u,v'\}\in E$, %
then it holds
\[\sigma_p(G,c)=\Theta(\max_{W\subseteq V}
\partial^2_\infty(G[W],c\res{E(W)})/\pnorm{c\res{E(W)}})\]
(cf. Corollary~\ref{splitcol}).  So we can indeed view parameter
$\sigma_p$ as a subgraph-monotone version of $\bnd^2_\infty$.
However, it is much more convenient to work with splitting sets
instead of strictly balanced $2$-colorings.

Our main theorem gives an upper bound on the min-max boundary
decomposition cost in terms of the $p$-splittability and the
\emph{maximum $c$-weighted degree} $\Delta_c:=\max_{v\in
  V}c(\delta(v))$. %
The time for computing $k$-colorings that achieve the bounds of the
theorem is almost the same as for computing cheap splitting sets in
the graph.

\begin{theorem}\label{maintheorem}
  Let $G$ be a graph with edge costs~$c$.  Then for all $k\in\N$ and $p>1$,
  \[
  \bnd^k_\infty(G,c)= O_p(\sigma_p\cdot (k^{-1/p} \cdot \pnorm c
  +\Delta_c)).
  \]
  Moreover, suppose one can compute splitting sets of cost at
  most $s\cdot \pnorm{c\res{W}}$ in time $t(|G[W]|)$ for all subgraphs
  $G[W]$, weights $w$ and splitting values $w^*$, where $t(n)\ge n$ is
  a linear function in $n$.
  Then, there exists an $O(t(|G|)\cdot\log k)$-time algorithm to
  compute strictly balanced $k$-colorings of $G$ with maximum boundary
  cost $O_p(s\cdot(k^{-1/p}\cdot \pnorm c +\Delta_c))$.
\end{theorem}
In the remainder of this section we draw a connection between the
above result and the more common notion of ``separator theorems'',
which were already mentioned in Section~\ref{sec:introduction}.  The
connection also allows us to formulate a result asserting the
tightness of our upper bound on the min-max boundary decomposition
cost.

In the following, we will assume that the considered instances $(G,c)$
consisting of a graph $G=(V,E)$ and edge costs $c\col E\ra \Rnn$ are
\emph{well-behaved}, i.e., the maximum degree $\Delta(G)$ is bounded
and the local fluctuation $c(\delta(v))/\min_{e\in \delta(v)} c(e)$ at
each vertex $v\in V$ is bounded.  In Appendix~\ref{lowersec} we
discuss to what extend this assumption is necessary.

Before formulating our results, we need to introduce a few notions
(cf. Appendix~\ref{lowersec}).  A subset $S\subseteq V$ of the
vertices is a \emph{balanced separator} with respect to weights $w\col
V\ra \Rnn$ if all components of $G[V\setminus S]$ have weight at most
$\frac{2}{3}\onorm{w}$.  The \emph{cost} of $S$ is the total cost
$\sum_{s\in S}c(\delta(s))$ of the edges incident to $S$.  Then, a
well-behaved instance $(G,c)$ has a $p$-separator theorem
(cf. Definition~\ref{def:sep}) if for all $W\subseteq V(G)$ and
weights $w\col W\ra \Rnn$ there exists $w$-balanced separators in
$G[W]$ of cost $O(\pnorm{c\res{W}})$.  Since well-behaved instances
with $p$-separator theorem have constant $p$-splittability
(cf. Lemma~\ref{balsep}), our main theorem translates to graphs with
separator theorems.

Roughly speaking, the theorem below shows that a well-behaved graph
class, which is closed under a reasonable ``similarity'' relation, has
small min-max boundary $k$-decomposition cost if and only if it has a
$p$-separator theorem for some $p>1$.
\begin{theorem}\label{thm:2}
  Let $(G,c)$ be a well-behaved instance.
  If $(G,c)$ has a $p$-separator theorem, then we have for all $ k\in
  \N$, \[\bnd_\infty^k(G,c)=O_p(\pnorm{c}/k^{1/p}+\inf{c}).\]

  If there exists a weight function $w\col V\ra \Rnn$ with $\inf w\le
  \onorm{w}/4$ such that all $w$-balanced separators of $G$ have cost
  $\alpha\cdot\pnorm{c}$, then for infinitely many $k\in \N$ there
  exist instances $(\tilde G,\tilde c)$ ``similar'' to $(G,c)$
  with \[\bnd_\infty^k(\tilde G,\tilde c)\gg \alpha \cdot
  (\pnorm{\tilde c}/k^{1/p}+\inf{\tilde c}).\]
\end{theorem}
In the theorem above, instance $(\tilde G,\tilde c)$ is similar to
$(G,c)$ in the sense that $\tilde G$ is the union of $k/4$ disjoint
isomorphic copies of $G$, and $\tilde c$ assumes for an edge in
$\tilde G$ the cost of the corresponding edge in $G$.
We remark that there are weights~$\tilde w$ for $\tilde G$ such that
every $k$-coloring $\chi$ of $\tilde G$ with roughly balanced weights,
i.e., $\max_i\tilde w(\ichi(i))\le 2\onorm {\tilde w}/k$, has average
boundary cost at least proportional to $\alpha\cdot(\pnorm{\tilde
  c}/k^{1/p}+\inf{\tilde c})$.  So we cannot expect better general
upper bounds than in Theorem~\ref{thm:2}, even if we relax the strict
balancedness constraint and consider the average instead of the
maximum boundary cost.

For the proof of Theorem~\ref{thm:2} we refer to
Appendix~\ref{lowersec}.
Notice that the first part is implied by Theorem~\ref{maintheorem} and
the fact that $\sigma_p$ is at most a constant for well-behaved graphs
with $p$-separator theorem. The second part follows from the
observation that a balanced separator of $G$ can be constructed from
each restriction $\chi\res{G\ho i}$ of a roughly balanced coloring
$\chi$ of $\tilde G$ to one of the copies of $G$, say $G\ho i$.

In Sections~3-5 we sketch a proof of Theorem~\ref{maintheorem}.
First, it is instructive to introduce further notation that allows to
formulate our results and proofs more easily.

\paragraph*{Further Notation.}
Let $\Phi\colon V\ra \Rnn$ be a non-negative function on the vertices
of a graph $G=(V,E)$.  We extend $\Phi$ on the power set of $V$
implicitly by the notation $\Phi(U) :=\sum_{u\in U} \Phi(u)$ for $U\in
2^V$.  So it is justified to call $\Phi$ a \emph{measure} on $V$.

Let $\chi$ be a $k$-coloring of $G$. The function
$\Phi\ichi\colon[k]\ra \Rnn$ with $(\Phi\ichi)(i):=\Phi (\ichi(i))$
maps each color to the \emph{$\Phi$-measure} or \emph{$\Phi$-weight}
of its color class.  So, $\inf{\Phi\ichi}$ is the \emph{maximum
  $\Phi$-measure} (of a color class) of~$\chi$.

For any non-negative discrete function $f$, we write $\avg f:=\onorm
f/k$ when the number $k$ of colors is understood. So,
$\avg{\Phi\ichi}$ is the \emph{average $\Phi$-measure} (of the color
classes) of $\chi$ and we have
$\avg{\Phi\ichi}=\onorm{\Phi}/k=\avg{\Phi}$.

Analogously, the function $\partial \ichi\colon [k]\ra \Rnn$ with
$(\partial \ichi) (i):=\partial(\ichi(i))$ maps each color to the
boundary cost of its color class.  So, $\inf{\partial\ichi}$ is the
\emph{maximum boundary cost} of~$\chi$ as in
Definition~\ref{def:strictcol} and $\avg{\partial\ichi}$ is the
\emph{average boundary cost} of coloring $\chi$.  Clearly,
$\avg{\partial\ichi}\le \inf{\partial\ichi}$.

For disjoint vertex sets $W_0,W_1\subseteq V$, we can combine two
$k$-colorings $\chi_0\col W_0\ra [k]$ and $\chi_1\col W_1\ra [k]$ into
a $k$-coloring of $W:=W_0\cup W_1$, by forming the \emph{direct sum}
$\chi_0\oplus\chi_1\col W\ra [k]$ with
$(\chi_0\oplus\chi_1)(i)=\chi_b(i)$ if $i\in W_b$.

\section{Multi-balanced colorings}
\label{multbalsec}\label{sec:multi-balanc-color}

In this section, we relax the strict constraints on the weights of the
color classes and consider (non-strictly) balanced colorings.

We say that a coloring $\chi$ of $G=(V,E)$ is \emph{(weakly) balanced}
with respect to a vertex measure $\Phi\colon V\ra \Rnn$ if
$\infnorm{\Phi\ichi}=O(\avg{\Phi}+\inf{\Phi}).$

Furthermore, we are interested in colorings that are not only balanced
with respect to a single measure $\Phi$ like the weights $w$, but with
respect to a constant number of measures $\Phi\ho 1,\ldots,\Phi\ho r$.
We call such colorings \emph{multi-balanced}.  We shall see that the
proof of Theorem~\ref{maintheorem} greatly benefits from the following
results about multi-balanced colorings.

The main result of this section is the following lemma, which provides
a bound on the minimum average boundary cost of multi-balanced
$k$-colorings.

\begin{lemma}[Multi-bal. Min-Avg. Boundary]\label{multballem}
  Let $G=(V,E)$ be a graph with edge costs $c\colon E\ra \Rnn$ and
  vertex measures $\Phi\ho 1,\ldots,\Phi\ho r$.  Then, there exists a
  $k$-coloring $\chi$ that is balanced with respect to $\Phi\ho1$
  through $\Phi\ho r$ and has average boundary cost at most
  proportional to $\sigma_p\cdot k^{-1/p}\cdot \pnorm{c}$.  More
  precisely, we have
  \begin{align*}
    \infnorm{\Phi\ho j\ichi} &=O_r(\avg{\Phi\ho j}+\inf{\Phi\ho
      j})\text{ for }j\in[r]
    \\
    \avg{\bnd\ichi}&=O_r( \sigma_p\cdot q\cdot k^{-1/p}\cdot
    \pnorm{c})
  \end{align*}
  with $1=\frac{1}{p}+\frac{1}{q}$.  One can compute such a coloring
  in time $O_r( t(|G|)\cdot \log k)$ with~$t$ as in
  Theorem~\ref{maintheorem}.
\end{lemma}

We remark that one can in fact guarantee $\inf{\Phi\ho 1\ichi}\le
3\avg{\Phi\ho 1}+O_r(\inf{ \Phi\ho 1})$ for the coloring obtained by
Lemma~\ref{multballem}.

Similar to \cite{spielman:minmax}, we observe that in the proof of
Lemma~\ref{multballem} the boundary cost function $\bnd:2^V\ra\Rnn$,
which assigns each subset $U$ of $V$ its boundary cost $\bnd U$, can
approximately be modeled as a vertex measure.  Hence, the boundary
cost of a coloring can be balanced by the methods developed for
Lemma~\ref{multballem}.  This idea yields the following proposition,
which makes up the first out three steps towards
Theorem~\ref{maintheorem}.

\begin{proposition}[Multi-bal. Min-Max Boundary]
  \label{multbalprop}
  Let $G$ be as in Lemma~\ref{multballem}.  Then, there exists a
  $k$-coloring $\chi$ which is balanced with respect to $\Phi\ho1$
  through $\Phi\ho r$ and has maximum boundary cost at most
  proportional to $\sigma_p\cdot(k^{-1/p}\cdot \pnorm{c}+\Delta_c)$,
  formally,
  \begin{align*}
    \inf{\Phi\ho j\ichi} &=O_r(\avg{\Phi\ho j}+\inf{\Phi\ho j})\text{
      for }j\in[r]
    \\
    \inf{\bnd\ichi}&=O_r(\sigma_p\cdot(q \cdot k^{-1/p}\cdot
    \pnorm{c}+\Delta_c)).
  \end{align*}
  One can compute such a coloring in time $O_r( t(|G|)\cdot \log k)$
  with $t$ as in Theorem~\ref{maintheorem}.
\end{proposition}

In order to show Lemma~\ref{multballem}, we need the following
auxiliary lemma, which itself can be viewed as a refined version of
Lemma~\ref{multballem} for the case $k=2$.

\begin{lemma}\label{multbalcut}
  Let $G$ be as in Lemma~\ref{multballem}.  Then, each vertex set
  $W\subseteq V$ can be $2$-colored such that the cost of the edges
  between the two color classes is
  at most $(2^r-1)\cdot \sigma_p(G,c)\cdot \pnorm {c\res W}$ and for
  all $j\in[r]$, the $\Phi\ho j$-measure of each color class does not
  exceed
  $\frac{3}{4}(\Phi\ho j(W)+2^{r-j}\inf{\Phi\ho j})$.  In particular,
  the $\Phi\ho 1$-measure of each color class is at most
  $\frac{1}{2}(\Phi\ho 1(W)+2^{r-1}\inf{\Phi\ho 1})$.

  Furthermore, such a $2$-coloring of $G[W]$ can be found in time
  $O_r(t(|G[W]|))$ where $t$ is as in Theorem~\ref{maintheorem}.
\end{lemma}
\begin{proof}
  By induction on $r\ge 1$.  First, graph $G$ is bisected into two
  parts $U_1$ and $U_2$ with respect to measure $\Phi\ho r$.  More
  precisely, from the definition of $\sigma_p(G,c)$ it follows that
  there exists a splitting set $U_1\subseteq W$ with cost at most
  $\bnd_W U_1\le \sigma_p(G,c)\cdot \pnorm{c\res W}$ such that
  \begin{equation}\label{eq:9}
    |\Phi\ho r(U_1)-\Phi\ho r(W)/2| \le \inf {\Phi\ho r}/2
  \end{equation}
  Let $U_2:=W\backslash U_1$ be the complement of $U_1$ within $W$.
  In the case $r=1$, the coloring $\chi\col W\ra\{1,2\}$ with
  $\chi\res{U_b}\equiv b$ fulfills all requirements of the lemma.

  Therefore, we may assume $r>1$.  By induction hypothesis, we can
  find $2$-colorings $\chi_1$ and $\chi_2$ of $G[U_1]$ and $G[U_2]$,
  that fulfill the conditions of the lemma for $\Phi\ho 1$ through
  $\Phi\ho {r-1}$, i.e., for $b\in\{1,2\}$ and $j\in[r-1]$,
  \begin{align}
    \inf{\Phi\ho j\ichi_b}&\le 3/4(\Phi\ho
    j(U_1)+2^{r-1-j}\inf{\Phi\ho j})
    \label{eq:i1}
    \\
    \inf{\bnd \ichi_b}&\le
    (2^{r-1}-1)\cdot\sigma_p(G,c)\cdot\pnorm{c\res{U_b}}\label{eq:i2}
  \end{align}
  Without loss of generality, we may assume that for $b\in\{1,2\}$,
  that \begin{equation}\label{eq:8} \Phi\ho r\ichi_b(b)\le
    \tfrac{1}{2}\Phi\ho r(U_b)
  \end{equation}
  Now, let $\chi\col W\ra \Rnn$ be the direct sum of $\chi_1$ and
  $\chi_2$, i.e., $\chi{\res {U_b}}=\chi_b$. Then
  \begin{align*}
    \Phi\ho r\ichi(b)
    &=\Phi\ho r\ichi_1(b)+\Phi\ho r\ichi_2(b)\\
    &\stackrel{\eqref{eq:8}}\le 1/2\cdot\Phi\ho r(U_b)+\Phi\ho r(W\backslash U_b)\\
    &\stackrel{\eqref{eq:9}}\le 1/4\cdot\Phi\ho r(W)+1/2\cdot\Phi\ho
    r(W)+3/4\cdot\inf{\Phi\ho r}
    \\
    & =3/4\cdot(\Phi\ho r(W)+2^{r-r}\inf{\Phi\ho r})
  \end{align*}
  We maintain for $j<r$ \[ \inf{\Phi\ho j\ichi}\le \sum_b \inf{\Phi\ho
    j\chi_b^{-1}} \stackrel{\eqref{eq:i1}}\le \tfrac{3}{4}(\Phi\ho
  j(W)+2^{r-j}\inf{\Phi\ho j})
  \]
  Similarly, we have $\inf{\bnd\ichi}\le \bnd_W U_1+\sum_b
  \inf{\bnd\chi_b^{-1}} \stackrel{\eqref{eq:i2}}\le
  (2^r-1)\cdot\sigma_p(G,c)\cdot\pnorm{c\res W}$ using the fact that
  $\pnorm{c\res {U_b}}\le \pnorm{c\res{W}}$ for $b\in \{1,2\}$.
  The additional guarantee \[ \inf{\Phi\ho 1\ichi}\le \sum_b
  \inf{\Phi\ho 1\chi_b^{-1}}\le \tfrac{1}{2}(\Phi\ho
  j(W)+2^{r-1}\inf{\Phi\ho 1})\] is also easily seen to be maintained.
\end{proof}

The proof of Lemma~\ref{multballem} is by induction on the number of
measures to be balanced.  Due to the length of the proof, we formulate
the induction step as a lemma of its own.  It states that given any
coloring $\chi$, one can efficiently compute a new coloring $\hat\chi$
which is balanced with respect to the measure $\Phi\ho 1$ such that
the maximum $\Phi\ho j$-measure ($1<j\le k$) and the average boundary
cost of the coloring increases by essentially at most a constant
factor.  So if $\chi$ was balanced with respect to measure $\Phi\ho 2$
to $\Phi\ho r$, then $\hat\chi$ is balanced with respect to measures
$\Phi\ho 1$ to $\Phi\ho r$.

\begin{lemma}\label{indstep}
  Let $G$ be as in Lemma~\ref{multballem} and let $\chi$ be an
  arbitrary $k$-coloring of $G$.

  Then, a $k$-coloring $\hat\chi$ of $G$ can be found in time
  $O_r(t(|G|)\log k)$ such that
  \begin{align*}
    \inf{\Phi\ho1\,\hichi} &= O_r(\avg{\Phi\ho 1}+\inf{\Phi \ho 1})
    \\
    \inf{\Phi\ho j \hichi} &= O_r(\inf{\Phi\ho j\ichi}+\inf{\Phi\ho
      j})
    \\
    \avg{\bnd\hichi} &=O_r(\inf{\bnd\ichi} +\mathcal B )
  \end{align*}
  with $\mathcal B=q\cdot k^{-1/p}\cdot\sigma_p\cdot\pnorm{c}$, and
  $t$ as in Theorem~\ref{maintheorem}.
\end{lemma}
Since the induction basis, $r=0$, for Lemma~\ref{multballem} is
trivial, it only remains to show Lemma~\ref{indstep} and
Proposition~\ref{multbalprop}.

\begin{proof}[Proof of Lemma~\ref{indstep}]
  During the construction of $\hat\chi$, for each color~$i\in[k]$ a
  \emph{tentative} color class~$\tent(i)\subseteq V$ is maintained.
  We start with $\tent(i)=\ichi(i)$ for each color~$i$.  The algorithm
  has the invariant:
  \begin{invariant}
    \label{inv:1}
    $\{\tent(i)\}_{i\in [k]}$ is a partition of $V$.
  \end{invariant}
  In fact, $\tent(i)$ will assume at most three different sets in the
  course of the algorithm for each color~$i$.  For convenience, let
  $\Psi:=\Phi\ho 1$.  According to the $\Psi$-weight of $\tent(i)$, we
  maintain a partition of the color set $[k]$.
  \begin{align*}
    \Light&=\{i\in [k]\mid \Psi(\tent(i))< \avg \Psi\}\\
    \Heavy&=\{i\in [k]\mid  \Psi(\tent(i))\ge 3\avg \Psi+2^r\inf\Psi \}\\
    \Medium&=[k]\setminus (Light \cup \Heavy)
  \end{align*}
  Each heavy color $i\in [k]$ will, in some iteration of our
  algorithm, be turned into a medium color by \textsc{Move}ing
  vertices from $\tent(i)$ to tentative color classes of light colors.

  Invariant~\ref{inv:1} and the definition of the partition
  $\{\Light,\Medium,\Heavy\}$ imply the claim below, asserting that we
  can assign at least two distinct light colors to every heavy color.
  \begin{claim}\label{clm:1}
    $|\Light|\ge 2|\Heavy|$.
  \end{claim}
  \begin{pf}
    We have $\avg\Psi|Medium|+3\avg\Psi|Heavy|\le \onorm\Psi= k\cdot
    \avg\Psi$.  Since $k=|\Light|+|\Medium|+|Heavy|$, we get
    $2|\Heavy|\le |Light|$. \smallqed
  \end{pf}
  We have another partition of $[k]$ into parts $\Untouch$, $\Pend$,
  and $\Finish$. Initially all colors are untouched.
  As the names suggest, we will have $\tent(i)=\ichi(i)$ for untouched
  colors and $\tent(i)=\hichi(i)$ for finished colors.  For pending
  colors, $\tent(i)$ is a common superset of both $\ichi(i)$ and
  $\hichi(i)$, and so we might be obliged to change tentative color
  classes of pending colors.

  At each point in the algorithm, the two color partitions that we
  maintain are related to each other in the following way.
  \begin{invariant}
    \label{inv:2}
    The following inclusions hold:
    \begin{align*}
      \Light\subseteq \Untouch &\quad\quad \Heavy\subseteq \Pend\\[1mm]
      \Finish&\subseteq \Medium
    \end{align*}
  \end{invariant}
  To set up the inclusion $\Heavy\subseteq \Pend$ in the beginning, we
  let each ``initially'' heavy color~$i$ with $\Psi\ichi(i)\ge
  3\avg\Psi+2^r\inf \Psi$ be $\Pend$.

  As indicated before, an iteration of our algorithm consists of
  \textsc{Move}ing vertices from one tentative color class to other
  tentative color classes.

  For each color $i\in[k]$, we will have a set $\Vin(i)$ of vertices
  \emph{in}coming to color~$i$ and a set $\Vout(i)$ of vertices
  \emph{out}going of color~$i$.  Similar to a network flow
  conservation law, we have
  \begin{equation}
    \label{eq:1}
    \ichi(i)\cup \Vin(i) = \hichi(i)\cup \Vout(i).
  \end{equation}
  None of $\Vin(i)$, $\hichi(i)$, or $\Vout(i)$ is a ``dynamic''
  sets. Once defined by the algorithm, the sets are not changed
  afterwards.  On the other hand, the two partitions of $[k]$ and the
  tentative color classes may change during the construction. As the
  names suggest, it is not possible that a color gets $\Untouch$ again
  after it was pending or even finished.

  We can now define the tentative color class $\tent(i)$ in terms of
  $\ichi(i)$, $\Vin(i)$, and $\hichi(i)$ depending on the current
  \emph{state} of color~$i$,
  \[
  \tent(i)=\begin{cases}
    \ichi(i)&\text{ if }i\in \Untouch,\\
    \ichi(i)\ \dot\cup\ \Vin(i)&\text{ if }i\in \Pend,\\
    \hichi(i)&\text{ if }i\in \Finish.
  \end{cases}
  \]
  Then, our algorithm consists of iterating the following procedure as
  long as there exist pending colors.  (Remember that we start with
  $\Pend=\{i\in[k]\mid \Psi\ichi(i)\ge 3\avg\Psi+2^r\inf \Psi\}$ and
  $\Finish=\emptyset$.)
  \begin{procedure}{Move}{color $i\in[k]$}
    \irem{Precondition: color $i$ is pending.}
  \item If $i\in \Medium$, \subitem then augment $\hat\chi$ such that
    $\hichi(i)=\tent(i)$, \subitem move $i$ from $\Pend$ to $\Finish$;
    return.  \irem{ If pending color $i$ is not medium, then $i\in
      \Heavy$ by Invariant~\ref{inv:2},\newline and so $|\Light|>2$ by
      Claim~\ref{clm:1}.}
  \item Choose distinct colors $x_1,x_2\in \Light$.
  \item Compute a splitting set $U$ in $G[X]$ of cost $\bnd_{X}U\le
    \sigma_p\cdot\pnorm{c\res{X}}$, where $X:=X(i):=tent(i)$ with $
    \avg\Psi\le \Psi(U)\le \avg\Psi+\inf\Psi.  $
  \item Find a $2$-coloring $\chi_0$ of $G[W]$ as in
    Lemma~\ref{multbalcut} where $W:=\Vout(i):=X\backslash U$.
  \item Augment coloring $\hat\chi$ such that $\hichi(i)=U$ and define
    $\Vin(x_b):=\ichi_0(b)$ for $b\in\{1,2\}$.
  \item Move color $i$ from $\Pend$ to $\Finish$ and colors $x_1,x_2$
    from $\Untouch$ to $\Pend$; return.
  \end{procedure}
  We observe that for $i,x_1,x_2$ as above, it holds
  \begin{equation}
    \Vout(i)=\Vin(x_1)\cup\Vin(x_2)\label{eq:2}
  \end{equation}
  The procedure $\Move$ maintains Invariant~\ref{inv:1}.  From the
  claim below it follows that also Invariant~\ref{inv:2} is
  maintained. The claim holds since both color classes of $\chi_0$
  have $\Psi$-weight at least $\avg \Psi$ for $i$ being heavy.
  \begin{claim}\label{clm:4}
    If procedure \Move is applied to a heavy color~$i$ and
    colors~$x_1,x_2$ are selected in step $(2.)$, then one has $i\in
    \Medium$ and $x_1,x_2\not\in \Light$ afterwards.
  \end{claim}
  \begin{pf}
    After the procedure, color $i$ is medium because
    $\Psi(tent(i))=\Psi(U)$ and $\avg\Psi\le\Psi(U)\le
    \avg\Psi+\inf\Psi$ by construction.  From Lemma~\ref{multbalcut},
    we get $\Psi \Vin( x_b)\ge 1/2\cdot( \Psi(W)-2^{r-1}\inf\Psi)$.
    By construction, we have $\Psi(W)=\Psi(X)-\Psi(U)$ and $\Psi(U)\le
    \avg\Psi+\inf \Psi$.  Since color $i$ was heavy when procedure
    $\Move$ was applied, we have $\Psi(X)\ge3\avg\Psi+2^r\inf\Psi$ and
    therefore
    \begin{align*}
      \Psi\Vin(x_b)&\ge 1/2\cdot( \Psi(X)-\Psi(U)-2^{r-1}\inf\Psi)\\
      &\ge 1/2\cdot(\Psi(X)-\avg\Psi-\inf\Psi-2^{r-1}\inf\Psi)\\
      &\ge 1/2\cdot(2\avg\Psi)=\avg\Psi.
    \end{align*}
    By construction, each color $x_b$ is $\Pend$ after $\Move(i)$ and
    hence $\Psi(\tent(x_b))\ge \Psi\Vin(x_b)\ge \avg\Psi$.  Thus,
    color $x_b$ cannot be $\Light$ anymore.\smallqed
  \end{pf}
  After termination of the algorithm, we have $\Pend=\emptyset$ and by
  Invariant~\ref{inv:2} also $\Heavy=\emptyset$.  So we obtain a
  $\Psi$-balanced coloring when we let the color classes of $\hat\chi$
  agree with the tentative color classes.
  We show next that our construction increased the maximum $\Phi\ho
  j$-measure by not more than a constant factor (essentially).

  The procedure \Move induces a tree-like structure on the set of
  colors.  More specifically, let $\mathcal F$ be the directed binary
  forest on nodes $[k]$ where a node $i\in[k]$ has children
  $x_1,x_2\in[k]$ if $\Vout(i)=\Vin(x_1)\cup \Vin(x_2)$.

  By Lemma~\ref{multbalcut} and identity~\eqref{eq:2} we have for each
  arc $(i,x)$ in $\F$, that $\Phi\ho j\Vin(x)\le 3/4\cdot \Phi\ho
  j\Vout(i) +O_r(\inf{\Phi\ho j})$.  This observation and
  identity~\eqref{eq:1} imply that the $\Phi\ho j$-weight of $\Vin(i)$
  decreases geometrically along the arcs of $\F$, i.e.,
  \begin{equation}
    \label{eq:3}
    \Phi\ho j\Vin(x)\le {\tfrac{3}{4}}\Phi\ho j\Vin(i)+O_r(\inf{\Phi\ho j\ichi}
    +\inf{\Phi\ho j})
  \end{equation}
  Since $\Vin(s)=\emptyset$ for each root $s$ of $\F$ and since the
  geometric series over $3/4$ is convergent, relation~\eqref{eq:3}
  implies $\Phi\ho j\Vin(i)=O_r(\inf{\Phi\ho j\ichi}+\inf{\Phi\ho
    j})$.  So the claim below holds for each color $i\in[k]$, because
  $\hichi(i)\subseteq X(i):=\ichi(i)\cup \Vin(i)$ by
  identity~\eqref{eq:1},
  \begin{claim}\label{clm:2}
    $\Phi\ho j\hichi(i)\le \max_i \Phi\ho j X(i) = 4\inf{\Phi\ho
      j\ichi}+O_r(\inf {\Phi\ho j})$.
  \end{claim}
  \begin{pf}
    By induction on the distance $h$ of color $i$ from a root $s$ in
    $\mathcal F$, we conclude from relation~\eqref{eq:3} that \[
    \Phi\ho j\Vin(i)\le(3/4)^h\Phi\ho j\Vin(s) + \sum_{l=1}^\infty
    (3/4)^l(\inf{\Phi\ho j\ichi}+O_r(\inf{\Phi\ho j})).\] Since
    $\Vin(s)=\emptyset$, we have
    \[\Phi\ho j\Vin(i)\le 3\inf{\Phi\ho j \ichi}+O_r(\inf{\Phi\ho
      j}).\]
    By identity~(\ref{eq:1}), we have $\Phi\ho j\hichi(i)\le\Phi\ho
    jX(i)= \Phi\ho j\Vin(i)+\inf{\Phi\ho j\ichi}$ and therefore
    $\inf{\Phi\ho j\hichi}\le \max_i \Phi\ho j X(i)\le 4\inf{\Phi\ho
      j\ichi}+O_r(\inf {\Phi\ho j})$.  \smallqed
  \end{pf}

  What remains is to estimate the average boundary cost of the
  coloring $\hat\chi$ in terms of $\avg{\bnd\ichi}$ and $\B$.
  By Lemma~\ref{multbalcut}, the cost of the edges cut by \Move$(i)$
  is at most proportional to $\sigma_p\cdot \pnorm{c\res {X(i)}}$.  So
  we have
  \begin{equation}\label{eq:4}
    \textstyle\avg{\bnd\hichi}
    \le \avg{\bnd\ichi}+O_r(\sigma_p\sum_{i=1}^k\pnorm{c\res {X(i)}}/k)
  \end{equation}
  To meet the requirements of the lemma, we need to show that
  $\sum_{i=1}^k\pnorm{c\res {X(i)}}=O_{r}(\B)$.  The idea is to
  consider first each component of $\F$ separately.  Let $C_s\subseteq
  [k]$ denote the tree component of $\F$ with root $s\in[k]$. We shall
  need a bound on the depth of $C_s$ in terms of the $\Psi$-weight of
  $\ichi(s)$.

  For a color~$i$, let $\exs(i):=\Psi X(i)-\avg \Psi$ be the amount by
  which the $\Psi$-weight of $\tent(i)$ exceeded the average
  $\Psi$-weight at the time when color~$i$ was pending.  Similar to
  relation~\eqref{eq:3}, $\exs(i)$ decreases geometrically along the
  arcs~$(i,x)$ of $\F$.  Since Lemma~\ref{multbalcut} gives stricter
  estimates for $\Phi\ho 1=\Psi$, the claim below follows in fact from
  identities~\eqref{eq:1} and \eqref{eq:2}.
  \begin{claim} \label{clm:5} $\exs(x)\le
    \frac{1}{2}\exs(i)+2^{r-2}\inf \Psi$
  \end{claim}
  \begin{pf}
    Identity~\eqref{eq:1} gives $\Psi X(x)= \Psi
    \ichi(x)+\Psi\Vin(x)$.  It holds $\Psi\ichi(x)\le \avg\Psi$,
    because color $x$ was $\Light$ in the beginning.  Since $(i,x)$ is
    a arc in $\F$, we get $\Psi\Vin(x)\le
    \Psi\Vout(i)/2+2^{r-2}\inf\Psi$ from Lemma~\ref{multbalcut} and
    identity~\eqref{eq:2}.  Again by identity~\eqref{eq:1} we have
    $\Psi\Vout(i)=\Psi X(i)-\Psi\hichi(i)$. By Claim~\ref{clm:4} it
    holds $\Psi \hichi(i)\ge \avg\Psi$ and thus $\Psi\Vout(i)\le
    \exs(i)$.  Now it follows
    \begin{align*}
      \exs(x)&=\Psi\Vin(x)+\Psi\ichi(x)-\avg\Psi\\
      &\le\Psi\Vout(i)/2+2^{r-2}\inf\Psi\\
      &\le\Psi(\exs(i)/2+2^{r-2}\inf\Psi.  \tag*{\scriptsize
        $\blacksquare$}
    \end{align*}
  \end{pf}
  By the definition of the color set $\Heavy$, every colors~$x$ with
  $\exs(x)\le 2\avg\Psi+2^r\inf\Psi$ is either $\Light$ or $\Medium$,
  and hence $x$ is a leaf in $\F$.  So it follows from
  Claim~\ref{clm:5} and a simple inductive argument, that the depth of
  component $C_s$ is at most logarithmic in the ratio of
  $\Psi\ichi(s)$ and $\avg \Psi$.
  \begin{claim}\label{clm:3}
    The depth $d_s$ of a $\mathcal F$-component with root $s$ is at
    most $\log (\Psi \ichi(s)/\avg\Psi)$.
  \end{claim}
  \begin{pf}
    From Claim~\ref{clm:5} it follows by induction on the depth $d$ of
    node $x\in C_s$, that the following relation holds:
    \[\exs(x)\le 2^{-d}\exs(s)+2^{r-1}\inf\Psi\le2^{-d}
    \Psi \ichi(s)+2^{r}\inf\Psi\] At depth $d_s-1$ it must hold
    $\Psi\ichi(s)/2^{d_s-1}+2^r\inf \Psi\ge \exs(x)>2\avg\Psi+2^r\inf
    \Psi$ for at least one $x\in C_s$. It follows $\Psi\ichi(s)\ge
    2^{d_s}\avg\Psi$ as claimed.\smallqed
  \end{pf}

  Claim~\ref{clm:3} implies that the running time of our algorithm is
  $O(t(|G|)\cdot\log k)$.  By Invariant~\ref{inv:1}, the vertex sets
  $X(i)$ are pairwise disjoint for all nodes~$i$ in the same level of
  $\F$, i.e., with the same distance from a root in $\F$.  By
  linearity of $t$, the total time for the colors in one level is
  $O(t(|G|))$.  So the total running time is $O(t(|G|)\cdot \log k)$
  by Claim~\ref{clm:3}.

  Since $\F[C_s]$ is a binary tree and the sets $X(i)$ are pairwise
  disjoint for nodes in the same level of $\F$, standard convexity
  arguments (H\"older's inequality) yield the claim below.
  \begin{claim}\label{clm:6}
    $\sum_{i\in C_s} \pnorm{c\res{X(i)}}\le
    \sum_{d=0}^{d_s}\pnorm{c\res{A_s}}\cdot 2^{d/q}\le 3q
    \pnorm{c\res{A_s}}\cdot 2^{d_s/q}$, where
    $\frac{1}{p}+\frac{1}{q}=1$ and $A_s:=\bigcup_{i\in C_s} X(i)$.
  \end{claim}
  \begin{pf}
    Let $L^d_s$ be the nodes in $C_s$ with distance $d$ from $s$.  As
    $|L^d_s|\le 2^d$, we have by H\"older's inequality
    \[
    \sum_{i\in L^d_s} \pnorm{c\res{X(i)}}\le \pnorm{\sum_{i\in
        L^d_s}c\res{X(i)}} \cdot 2^{d/q}.
    \]
    Since the sets $X(i)\subseteq A_s$ are pairwise disjoint for $i\in
    L^d_s$, it holds $\sum_{i\in L^d_s} c\res{X(i)}\le c\res{A_s}$ and
    thus $\sum_{i\in L^d_s} \pnorm{c\res{X(i)}}\le \pnorm{c\res{A_s}}
    \cdot 2^{d/q}$.  So we can conclude
    \begin{align*}
      \sum_{i\in C_s}\pnorm{c\res{X(i)}}&\le \sum_{0\le d\le d_s}
      \pnorm{c\res{A_s}} \cdot 2^{d/q} \le \pnorm{c\res{A_s}} \cdot
      \tfrac{2^{1/q}}{2^{1/q}-1}\cdot 2^{d_s/q} \le
      3q\pnorm{c\res{A_s}} \cdot 2^{d_s/q}
    \end{align*}
    using $1/(2^{1/q}-1)\ge q/\ln 2$ and $2^{1/q}/\ln 2\le
    3$.\smallqed
  \end{pf}
  Using the bound on $d_s$ from Claim~\ref{clm:3}, and the bound on
  $\sum_{i\in C_s} \pnorm{c\res{X(i)}}$ from Claim~\ref{clm:6}, we
  arrive with H\"older's inequality at:
  \begin{claim}
    $\sum_{i=1}^k \pnorm{c\res{X(i)}}\le 3q \pnorm{c}\cdot
    k^{1/q}=3k\cdot \B$.
  \end{claim}
  \begin{pf}
    By Claim~\ref{clm:3} we have $d_s\le \Psi\ichi(s)/\avg\Psi\le
    \Psi(A_s)/\avg\Psi$.  Now it follows from Claim~\ref{clm:6} and
    H\"older's Inequality,
    \[
    \sum_{i=1}^k \pnorm{c\res{X(i)}} \le \sum_{s} 3q
    \pnorm{c\res{A_s}}(\tfrac{\Psi (A_s)}{\avg \Psi})^{1/q} \le 3q
    \Big\|\sum\nolimits_s c\res{A_s}\Big\|_p\cdot \Big(\sum\nolimits_s
    \tfrac{\Psi (A_s)}{\avg \Psi}\Big)^{1/q}.
    \]
    Since $\{A_s\}_s$ form a partition of the vertex set $V$, we
    conclude
    \begin{equation}
      \sum_{i=1}^k \pnorm{c\res{X(i)}}\le 3q \pnorm{c}\cdot (\onorm\Psi/\avg \Psi)^{1/q}
      =3q\pnorm c\cdot k^{1/q}.\tag*{\scriptsize$\blacksquare$}
    \end{equation}
  \end{pf}
  So by relation~(\ref{eq:4}), $\avg{\bnd\hichi}$ is at most the
  average boundary cost of the original coloring $\chi$ plus
  $O_r(q\cdot\sigma_p\cdot\pnorm c /k^{1/p})$.
\end{proof}

In the remainder of the section, we prove
Proposition~\ref{multbalprop}, which is our strongest result about
multi-balanced colorings.  The idea is to start with a coloring as
obtained by Lemma~\ref{multballem}, and then to balance the boundary
costs of the coloring using essentially the algorithm of
Lemma~\ref{indstep}.  The hope is that boundary costs of the color
classes behave in the algorithm approximately like vertex measures.
So, we should ensure that a single $\Move$-step does not break the
bound of Proposition~\ref{multbalprop} on the maximum boundary
costs. For this reason, we balance the coloring beforehand with
respect to the following measure.
\begin{definition}[Splitting Cost Measure]\label{def:splitmeasure}
  The \emph{(p-)splitting cost measure} of graph $G=(V,E)$ with
  respect to edge costs $c$ is defined by
  \[
  \pi\colon V\ra \Rnn,\quad \pi(v):= \sigma^p_p \sum_{e\in
    \delta(v)}c^p_e/2.
  \]
  For all vertex sets $W\subseteq V$, one has
  $\sigma_p\pnorm{c\res{W}}\le (\pi(W))^{1/p}=:\pi^{1/p}(W)$.  So
  there exist splitting sets in $G[W]$ of boundary cost at most
  $\pi^{1/p}(W)$, even if weights and splitting value are worst
  possible.  Thus, we call $\pi^{1/p}(W)$ the \emph{splitting cost} of
  $W$.
\end{definition}
Since $\onorm{\pi}^{1/p}=\sigma_p\pnorm{c}$ and $\inf \pi^{1/p}\le
\sigma_p\Delta_c$, we have
\begin{equation}\label{eq:5}
  (\onorm{\pi}/k+\inf\pi)^{1/p}\le \sigma_p(q\cdot k^{-1/p}\cdot\pnorm{c}+\Delta_c)=:\B'
\end{equation}
and hence color classes of $\pi$-balanced colorings can be split at
cost $O(B')$.  So if we start with a $\pi$-balanced coloring and
maintain the $\pi$-balancedness, then a single iteration of procedure
\Move from Lemma~\ref{indstep} cannot break the bound on the maximum
boundary cost from Proposition~\ref{multbalprop}.

\begin{proof}[Proof of Proposition~\ref{multbalprop}]
  We may assume that only the measures $\Phi\ho 3$ through $\Phi\ho r$
  are arbitrary and that the measures $\Phi\ho 1$ and $\Phi\ho 2$ can
  be defined by ourselves.  This assumption does not weaken the
  statement of the proposition since the statement is invariant under
  renaming and adding (constantly many) ``new'' measures.

  We define $\Phi\ho2$ to be the $p$-splitting cost measure $\pi$ of
  $G$ with respect to $c$. Let $\chi$ be a coloring that is balanced
  with respect to $\Phi\ho 2$ through $\Phi\ho r$ and has average
  boundary cost at most proportional to $\sigma_p\cdot k^{-1/p}\cdot
  \pnorm c$.  By Lemma~\ref{multballem} such a coloring exists and can
  be obtained efficiently.

  Consider the following measure that accounts for edges not running
  within a single color class of $\chi$
  \[
  \Psi(v):= c(\{uv\in E\mid \chi(u)\neq \chi(v)\}).
  \]
  Note that $\inf{\bnd\ichi}= \inf{\Psi\ichi}$,
  $\avg\Psi=\avg{\bnd\ichi}$, and $\inf \Psi\le \Delta_c$.  So if
  $\chi$ was $\Psi$-balanced, then the maximum boundary cost of $\chi$
  would already be as required by the proposition.

  We use Lemma~\ref{indstep} to establish $\Psi$-balancedness.
  However, there is a small twist.
  Instead of instantiating the lemma for $r$ measures, we instantiate
  it for $r+1$ measures, namely $\Phi\ho 1,\ldots,\Phi\ho {r+1}$,
  where $\Phi\ho1:=\Psi$, $\Phi\ho 2:=\pi$, and $\Phi\ho{r+1}$ is
  defined later.  Then, let $\hat\chi$ be the coloring obtained by the
  algorithm of Lemma~\ref{indstep} for measures $\Phi\ho 1$ through
  $\Phi\ho {r+1}$.  So $\hat\chi$
  is balanced with respect to $\Phi\ho 1$ through $\Phi\ho r$ and has
  average boundary cost at most proportional to $\sigma_r\cdot
  k^{-1/p}\cdot \pnorm{c}$.

  The idea is now that for vertex sets $U\subseteq V$, the $\Phi\ho
  1$-weight of $U$ approximates the boundary cost of $U$ in the graph
  induced by the $\chi$-bichromatic edges, and the
  $\Phi\ho{r+1}$-weight of $U$ shall approximate the boundary cost of
  $U$ within the monochromatic edges.  So taking these two measure
  together should yield an approximation of the ``real'' boundary cost
  of $U$.

  For the following proof, we define all symbols as in the proof of
  Lemma~\ref{multballem}.  In addition, let $E':=\{e\in E\ \mid \
  |\chi(e)|=1\}$ be the set of edges that run within one color class
  of $\chi$.  Note that the measure $\Psi$ accounts for all edges
  besides $E'$.  More specifically, we have for all $U\subseteq V$,
  \begin{equation}
    \partial U \le \Psi U+\bnd'(U)\label{eq:15}
  \end{equation}
  where $\bnd'U:=c(\delta(U)\cap E')$ is the cost of the
  $\chi$-monochromatic boundary edges.
 
  Since $\hat\chi$ is balanced with respect to $\Psi$, we know
  $\inf{\Psi\hichi}=O_r(\avg\Psi+\inf\Psi)=O_r(\B')$,
  Hence the estimate $\inf{\partial'\hichi}=O_r(\B')$ will imply the
  desired bound $\inf{\bnd\hichi}\le
  \inf{\Psi\hichi}+\inf{\bnd'\hichi}=O_r(\B')$.

  For deriving $\inf{\partial'\hichi}=O_r(\mathcal B')$, we need the
  following two technical claims.

  \setcounter{claim}{7}

  Stronger than the $\pi$-balancedness of $\hat\chi$, the claim below
  holds -- indeed implied by the discussion for Claim~\ref{clm:2}.
  \begin{claim}\label{clm:7}
    $\pi^{1/p}(X(i))\ll_r(\avg \pi+\inf
    \pi)^{1/p}\stackrel{\eqref{eq:5}}{=}O_r(\B')$.
  \end{claim}
  \begin{pf}
    By Claim~\ref{clm:2}, we have $\pi(X(i))\le
    4\inf{\pi\ichi}+O_r(\inf{\pi})$.  The balancedness of $\chi$ with
    respect to $\pi$ implies $\inf{\pi\ichi}=O_r(\avg \pi
    +\inf\pi)$. \smallqed
  \end{pf}
  The next claim is a consequence of identity~\eqref{eq:1}, and the
  fact that the edges between parts $\hichi(i)$ and $\Vout(i)$ have
  cost at most $\pi^{1/p}(X(i))=O_r(\B')$.
  (Note that we can assume $\hichi(i)\neq\ichi(i)$, since otherwise
  $\bnd'\hichi(i)=\bnd'\ichi(i)=0$.)
  \begin{claim}\label{clm:9}
    If $\hichi(i)\neq \ichi(i)$ then $\partial' \hichi(i)\le
    \bnd'\Vin(i)+O_r(\B')$.
  \end{claim}
  \begin{pf}
    If the $i$-th color classes in $\chi$ and $\hat\chi$ differ, then
    $\Vin(i)$ was defined at some point of the algorithm.  Since
    $\hichi(i)\subseteq X(i)=\ichi(i)\cup\Vin(i)$ by
    identity~\eqref{eq:1}, every boundary edge of $\hichi(i)$ either
    crosses $X(i)$ or completely runs within $X(i)$. Thus
    $\bnd'\hichi(i)\le \bnd_{X(i)}\hichi(i)+\bnd'X(i)$.  By
    construction, the boundary cost of $\hichi(i)$ in $G[X(i)]$ is at
    most $\pi^{1/p}(X(i))=O_r(\B')$.  Since
    $\bnd'$ vanishes on $\ichi(i)$ as all its boundary edges are
    bichromatic in $\chi$, we have $\bnd'X(i)\le
    \bnd'\ichi(i)+\bnd'\Vin(i) =\bnd'\Vin(i)$.\smallqed
  \end{pf}
  For the bound $\inf{\bnd'\hichi}=O_r(\B')$, it is now sufficient to
  show $\bnd'\Vin(i)=O_r(\B')$.  
  The idea is to achieve a situation for $\bnd'\Vin(i)$ similar to
  relation~\eqref{eq:3}, i.e., we want $\bnd'\Vin(i)$ to decrease
  geometrically along the arcs of $\F$.  In fact, nothing else remains
  to show:
  \begin{claim}\label{clm:8}
    If $\bnd'\Vin(i)$ decreases geometrically along the arcs of $\F$,
    then the maximum boundary cost of $\hat\chi$ with respect to both
    $\bnd'$ and $\bnd$ is in $O_r(\B')$.
  \end{claim}
  \begin{pf}
    If it holds $\bnd'\Vin(x)\le z\cdot \bnd'\Vin(i)+O_r(\B')$ for
    some fixed constant $z<1$ and all arcs $(i,x)$ of $\F$, then we
    have $\bnd'\Vin(i)=O_r(\B')$ and by Claim~\ref{clm:9} also
    $\bnd'\hichi(i)=O_r(\B')$ for all colors $i\in [k]$.
    Then, it follows from relation~\eqref{eq:15} that the maximum
    boundary cost of $\hichi$ bounded by $\inf{\Psi\hichi}+O_r(\B')$.
    By construction, $\hichi$ is balanced with respect to $\Psi$ and
    thus $\inf{\Psi\hichi}\ll_r\avg\Psi+\inf\Psi \ll_r
    \inf{\bnd\ichi}+\Delta_c = O_r(\B')$.  Hence we get
    $\inf{\bnd\hichi}=O_r(\B')$.  \smallqed
  \end{pf}
 
  We choose the measure $\Phi\ho{r+1}$ in the following way to ensure
  the assumption of Claim~\ref{clm:8}.  At the time when \Move is
  applied to color $i\in [k]$, let $\Phi\ho{r+1}(v)$ be the cost of
  the edges from $\delta(\tV(i))\cap E'$ that are incident to $v\in V$
  Formally, $\Phi\ho{r+1}(v):=c(\delta(v)\cap\delta(\tV(i))\cap E')$
  for $v\in \Vin(i)$.  For convenience, we set $\Phi\ho {r+1}(v)=0$
  for vertices outside of $\Vin(i)$.  We have
  $\Phi\ho{r+1}\Vin(i)=\bnd'\Vin(i)$ and for all $U\subseteq V$, in
  particular $U=\Vin(i)$, it holds
  \begin{equation}
    \bnd'U
    \le \Phi\ho{r+1}(U)+\bnd_{X(i)}U\label{eq55}
  \end{equation}
  by identity~\eqref{eq:1} and the fact $\bnd'\ichi(i)=0$.  So
  $\Phi\ho{r+1}$ is a good approximation of $\bnd'$, at least for sets
  with small boundary cost in $G[X(i)]$.

  For an arc $(i,x)$ of $\F$, it follows from Claim~\ref{clm:7} and
  Lemma~\ref{multbalcut} that the boundary cost of $\Vin(x)$ in
  $G[X(i)]$ is at most proportional to $\pi^{1/p}(X(i))=O_r(\B')$.
  Lemma~\ref{multbalcut} guarantees $\Phi\ho{r+1}\Vin(x)\le
  3/4\cdot\Phi\ho{r+1}\Vout(i)+O_r(\inf{\Phi\ho{r+1}})$.  Then
  relation~\eqref{eq55} and the fact $\Phi\ho{r+1}
  \Vout(i)\le\bnd'\Vin(i)$ finally show:
  \begin{claim}
    $\bnd'\Vin(x)\le \frac{3}{4}\bnd'\Vin(i)+O_r(\B')$.
  \end{claim}
  \begin{pf}
    Using $\bnd_{X(i)} \Vin(x)=O_r(\B')$ and $\Phi\ho{r+1}\Vin(x)\le
    3/4\cdot\Phi\ho{r+1}\Vout(i) +O_r(\inf{\Phi\ho{r+1}})$, we get
    from relation~\eqref{eq55}: $\bnd'\Vin(x)\le 3/4\cdot
    \Phi\ho{r+1}\Vout(i)+O_r(\B'+\inf{\Phi\ho{r+1}})$.  Clearly,
    $\inf{\Phi\ho{r+1}}\le \B'$.  Since
    $\onorm{\Phi\ho{r+1}}=\bnd'\Vin(i)$ when $\Move$ is applied to
    color~$i$, we have $\Phi\ho{r+1}\Vout(i)\le \bnd'\Vin(i)$ and
    therefore $\bnd'\Vin(x)\le 3/4\cdot \bnd'\Vin(i)+O_r(\B')$, as
    required.  \smallqed
  \end{pf}
  From Claim~\ref{clm:8} it follows now that $\hat\chi$ fulfills all
  requirements of the proposition.
\end{proof}

\section{Improving balancedness at no cost}\label{strictsec}

In this section we show how weakly balanced colorings can be
transformed into strictly balanced colorings while maintaining the
bounds on the maximum boundary cost claimed by
Theorem~\ref{maintheorem}.  Together with
Proposition~\ref{multbalprop}, this result shall imply
Theorem~\ref{maintheorem}.

We proceed in two steps.  First we obtain a similar result about
colorings with only slightly relaxed constraints on the weights.
A $k$-coloring is called \emph{almost strictly balanced} with respect
to weights $w$ if the weight of each color classes differs from the
average weight by at~most~$2\inf{w}$.

\begin{proposition}\label{shrinkprop}
  Let $k\in\mathbb N$ and $G=(V,E)$ be a graph with edge costs $c$ and
  vertex weights $w$.

  Then any $w$-balanced $k$-coloring $\chi$ of $G$ can be transformed
  into an almost strictly balanced $k$-coloring $\hat\chi$ without
  increasing the maximum boundary cost or splitting cost %
  by more than a constant factor, essentially.  More precisely,
  \begin{align*}
    \inf{\pi\hichi}&=O_p(\inf{\pi\ichi})\\
    \inf{\bnd\hichi}&=O_p(\inf{\bnd\ichi}+\inf{\pi\ichi}^{1/p})
  \end{align*}
  where $\pi$ is the $p$-splitting cost measure of $G$
  (cf. Definition~\ref{def:splitmeasure}).

  The coloring $\hat\chi$ can be obtained from $\chi$ in time
  $O(t(|G|))$ with $t$ is as in Theorem~\ref{maintheorem}.
\end{proposition}

As soon as we have an almost strictly balanced coloring its is easy to
obtain a desired strictly balanced coloring.  The idea is to reduce
the weight of each color class below $\avg{w}$ by cutting off parts
(vertex sets) of weight about $\inf{w}$. These parts are then
redistributed among the color classes by a greedy bin-packing
procedure.  Since we started with an almost strictly balanced
coloring, the above procedure alters each color class at most a
constant number of times. So we get the proposition below. For a
detailed proof we refer to Appendix~A.2.%

\begin{proposition}\label{greedyprop}
  Let $k$ and $G$ be as in Proposition~\ref{shrinkprop}.  Then any
  almost balanced $k$-coloring $\chi$ can be turned into a strictly
  balanced $k$-coloring with
  \[
  \inf{\bnd\hichi}=O_p(\inf{\bnd\ichi}+\inf{\pi\ichi}^{1/p}+\Delta_c).
  \]
  The coloring $\hat\chi$ can be obtained from $\chi$ in time
  $O(t(|G|)\log k)$, where $t$ is as in Theorem~\ref{maintheorem}.
\end{proposition}

Theorem~\ref{maintheorem} is implied by the conjunction of
Propositions~\ref{multbalprop}, \ref{shrinkprop}, and
\ref{greedyprop}.
\begin{proof}[Proof of Theorem~\ref{maintheorem}]
  If we apply Proposition~\ref{multbalprop} with $\Phi\ho 1:=w$ and
  $\Phi\ho 2:=\pi$, then we obtain a $w$-balanced coloring $\chi_1$
  such that both the maximum splitting cost $\inf{\pi\ichi_1}^{1/p}$
  and the maximum boundary cost $\inf{\bnd\ichi_1}$ of $\chi_1$ are at
  most proportional to $ \sigma_p\cdot
  (k^{-1/p}\cdot\pnorm{c}+\Delta_c)$.

  By Proposition~\ref{shrinkprop}, we can transform $\chi_1$ into an
  almost strictly $w$-balanced coloring $\chi_2$ with maximum
  splitting cost and maximum boundary cost fulfilling the same bounds
  as before.

  Finally, Proposition~\ref{greedyprop} yields a strictly $w$-balanced
  coloring $\chi_3$ of $G$ with maximum boundary cost $O_p(
  \sigma_p\cdot (k^{-1/p}\pnorm{c}+\Delta_c)$.  Hence it holds
  $\bnd^k_\infty(G,c)= O_p( \sigma_p\cdot
  (k^{-1/p}\pnorm{c}+\Delta_c)$.
\end{proof}
In the remainder of this section we sketch a proof of
Proposition~\ref{shrinkprop}.

We aim for a recursive algorithm that computes an almost strictly
balanced coloring from a weakly balanced $k$-coloring $\chi$, with
$\inf{w\ichi}\le M \avg w$ for some constant $M$.  First, a so called
\emph{shrinking procedure} computes from $\chi$ two colorings
$\chi_0\col V_0\ra [k]$ and $\chi_1\col V_1\ra [k]$ of disjoint vertex
subsets $V_0$ and $V_1$ with $V_0\cup V_1=V$.  The coloring $\chi_0$
shall be almost strictly balanced and $\chi_1$ is weakly balanced with
$\inf{w\ichi_1}\le M\avg{w\res{V_1}}$.  Most importantly, the maximum
splitting cost and the maximum boundary cost decrease geometrically
when going from coloring $\chi$ to the ``shrunken'' coloring $\chi_1$.

From coloring $\chi_1$, we recursively compute an almost strictly
balanced coloring~$\hat\chi_1$ of $V_1$. Since now both $\chi_0$ and
$\hat\chi_1$ are almost strictly balanced, the weight of each color
class in the direct sum ${\chi_0}\oplus{\hat\chi_1}\col V\ra[k]$
differs from the average weight by at most~$4\inf w$. So coloring
$\chi_0$ need to be changed only slightly to obtain a coloring
$\tilde\chi_0$ such that the direct sum $\tilde\chi_0\oplus
\hat\chi_1$ indeed is the desired almost strictly balanced coloring
$\hat \chi$ of $V$.

In order to ensure that the boundary costs do not accumulate in the
recursive calls, we need precise bounds on the maximum boundary cost
and maximum splitting cost of the colorings $\chi_0$ and $\chi_1$.
The following definition captures these (technical) requirements.
Roughly speaking, one wants that the maximum boundary cost
$\inf{\bnd\ichi_0}$ and the maximum splitting cost $\inf{\pi\ichi_0}$
of coloring $\chi_0$ are at most proportional to the respective costs
of the original coloring $\chi$.  The costs $\inf{\bnd{\ichi_1}}$ and
$\inf{\pi{\ichi_1}}$ of the coloring $\chi_1$ should be geometrically
less than the respective costs in $\chi$.  In order to ensure that our
recursive algorithm runs in linear time, we require that the size
$|G[V_1]|$ of the graph induced by $V_1$ is only a constant fraction
of $|G|$.

\begin{definition}[Shrinking Procedure]\label{def:shrink}
  For $\epsilon>0$ and $M:=1/\epsilon^5$, let $P$ be a procedure that
  transforms any weakly balanced $k$-coloring $\chi$ of a vertex set
  $W\subseteq V$ with \[ \inf{w\ichi}\le M\cdot \avg {w\res{W}} \quad
  \text{ and }\quad\inf{w}\le \epsilon^{5}\cdot\avg{ w\res W}
  \]
  into two $k$-colorings $\chi_0$ and $\chi_1$ of disjoint sets $W_0$
  and $W_1$ with $W_0\cup W_1=W$.

  Then procedure $P$ is called \emph{$\epsilon$-shrinking} if
  \begin{enumerate}[a)]
  \item coloring $\chi_0$ is almost strictly balanced with
    $w\ichi_0(i)-\epsilon\avg{w\res W}\in[0,\inf w]$, and it holds
    $\inf{\pi\ichi_0}=O_M(\inf{\pi\ichi})$, and also
    $\inf{\bnd\ichi_0}=O_M(\inf{\bnd\ichi}+\inf{\pi\ichi}^{1/p})$,
  \item coloring $\chi_1$ is weakly balanced with $\inf{w\ichi_1}\le
    M\avg {w\res{ W_1}}$, and it holds $\inf{\pi\ichi_1}\le
    (1-\epsilon^{10})\inf{\pi\ichi}$, and also $\inf{\bnd\ichi_1}\le
    (1-\epsilon^{10})\inf{\bnd\ichi}+O_M(\inf{\pi\ichi}^{1/p})$,
  \item it holds $|G[W_1]|\le (1-\epsilon^{10})|G[W]|$.
  \end{enumerate}
  Notice that $\bnd \ichi_b$ and $\bnd\ichi$ refer to the boundary
  costs of the respective color classes with respect to the host graph
  $G$ (as opposed to $G[W_b]$ or $G[W]$).
\end{definition}

The definition above makes only sense if there are efficient shrinking
procedures.
\begin{lemma}\label{shrinkproc}
  For sufficiently small $\epsilon>0$, there exist
  $\epsilon$-shrinking procedures that run in time proportional to
  $t(|G[W]|)$ when applied to a balanced coloring of $G[W]$.
\end{lemma}

We can think of our algorithm as a divide-and-conquer algorithm.  Then
the shrinking procedure divides the problem into two subproblems,
where the subproblem corresponding to the coloring $\chi_0$ is
trivial, since $\chi_0$ is already almost strictly balanced, and the
subproblem for coloring $\chi_1$ is of the same type as for the input
coloring $\chi$ but has reduced complexity, in the sense that the
maximum splitting and boundary costs decreased geometrically. We do
not know yet how the conquer-phase works, i.e., how to construct a
solution for the original problem.

Suppose we obtained an almost strictly balanced coloring $\hat\chi_1$
from the recursive call for $\chi_1$. We want to transform the
coloring $\chi_0$ into a coloring $\tilde\chi_0$ such that the direct
sum of $\tilde\chi_0$ and $\chi_1$ is almost strictly balanced. The
idea is to uncolor parts of the color classes in $\chi_0$ until the
direct sum with $\chi_1$ has maximum weight at most~$\avg w$.  The
weight of each part shall be between $\inf w$ and $2\inf w$.  Then
these parts are redistributed among the %
color classes %
by a greedy bin-packing procedure, so that the direct sum
$\tilde\chi_0\oplus\hat\chi_1$ is almost strictly balanced.

With the technical requirements of the lemma below, a straight-forward
analysis shows that each color class is changed only constantly often
in the conquer-phase and thus the the maximum splitting cost measure
or the maximum boundary cost increased by no more than a constant
factor (essentially).
\begin{lemma}[Conquer-Phase]\label{conquerlem}
  Let $\chi_0$ and $\hat\chi_1$ be two $k$-colorings of disjoint sets
  $W_0$ and $W_1$ with $W_0\cup W_1=W$.  Suppose that $w\hichi_1(i)\le
  \avg{w\res{W}}-\inf w$ for each color $i\in [k]$.

  If both $w\ichi_0(i)=\avg{w\res{W_0}}+O(\inf w)$ and
  $w\hichi_1(i)=\avg{w \res{W_1}}+O(\inf w)$ for all colors $i\in[k]$,
  then $\chi_0$ can be transformed into a coloring $\tilde\chi_0$ such
  that the direct sum $\hat\chi=\tilde\chi_0 \oplus \hat\chi_1$ is
  almost strictly balanced.

  Neither the maximum splitting cost nor the maximum boundary cost is
  increased by more than essentially a constant factor; more
  precisely, $\inf{\pi\tichi_0}=O(\inf{\pi\ichi_0})$ and
  $\inf{\bnd\tichi_0}=O(\inf{\bnd\ichi_0}+\inf{\pi\ichi_0}^{1/p})$.

  The coloring $\tilde\chi_0$ can be obtained in time $O(t(|G|))$.
\end{lemma}
Assuming Lemma~\ref{shrinkproc} and Lemma~\ref{conquerlem}, a
straight-forward analysis of the described ``shrink-and-conquer''
algorithm proves Proposition~\ref{shrinkprop}.
(Lemma~\ref{conquerlem} is also used to handle the base case of the
algorithm, i.e., for $\inf w\ge \epsilon^5\avg {w\res W}$ when the
$\epsilon$-shrinking procedure cannot be applied.)

For proofs of the assumed lemmas~\ref{shrinkproc} and \ref{conquerlem}
we refer to Section~\ref{sec:shrink} and Appendix~A.2, %
respectively.

\begin{proof}[Proof of Proposition~\ref{shrinkprop}]

  We show by induction on the cardinality of $W$ that the coloring
  $\chi$ can be transformed into an almost strictly balanced coloring
  $\hat\chi$ with $\inf{\pi\hichi}\le C_1\inf{\pi\ichi}$ and
  $\inf{\bnd\hichi}\le C_1\inf{\bnd\ichi}+C_2\inf{\pi\ichi}^{1/p}$ for
  appropriately chosen constants $C_1,C_2$.  Let $P$ be an
  $\epsilon$-shrinking procedure from Lemma~\ref{shrinkproc} for some
  sufficiently small absolute constant $\epsilon>0$.  Notice that
  $M:=1/\epsilon^5$ is also an absolute constant.

  The base case is $\inf{w}>\epsilon^{5}\avg{w\res{W}}$.  Since $\chi$
  is weakly balanced, the maximum weight of $\chi$ is at most
  $M\cdot\avg{w\res W}\le M^2\inf w$.  Hence we can apply
  Lemma~\ref{conquerlem} (with $W_0=W$ and $W_1=\emptyset$) to obtain
  an almost strictly balanced coloring $\tilde\chi_0$.  The coloring
  $\hat\chi:=\tilde\chi_0$ satisfies the requirements of the
  proposition.

  We may now assume $\inf{w}\le \epsilon^{5}\avg{w\res{W}}$ and
  $\avg{w\res{W}}>0$.  Then we can apply our $\epsilon$-shrinking
  procedure $P$ to obtain colorings $\chi_0$ and $\chi_1$ of disjoint
  vertex sets $W_0$ and $W_1$.  By induction hypothesis, $\chi_1$ can
  be transformed into an almost strictly balanced coloring
  $\hat\chi_1$ with $\inf{\pi\hichi_1}\le C_1\inf{\pi\ichi_1}$ and
  $\inf{\bnd\hichi_1}\le
  C_1\inf{\bnd\ichi_1}+C_2\inf{\pi\ichi_1}^{1/p}$.

  The maximum weight $\inf{w\hichi_1}$ is bounded by
  \[\avg{w\res{W_1}}+2\inf w
  \le (1-\epsilon+3\epsilon^5)\avg{w\res W}-\inf w\] which is less
  than the upper bound $\avg{w\res W}-\inf w$ required by
  Lemma~\ref{conquerlem}, for sufficiently small $\epsilon$.  Thus, we
  can apply Lemma~\ref{conquerlem} to obtain a coloring $\tilde\chi_0$
  such that $\tilde\chi_0\oplus\hat\chi_1$ is almost strictly
  balanced.

  The maximum $\pi$-weight of $\hat\chi:=\tilde\chi_0\oplus\hat\chi_1$
  is at most $\inf{\pi\hichi_1}+\inf{\pi\tichi_0}\le
  C_1\inf{\pi\ichi_1}+O(\inf{\pi\ichi})$.  Since $P$ is
  $\epsilon$-shrinking, it holds $\inf{\pi\ichi_1}\le z\inf{\pi\ichi}$
  for $z:=(1-\epsilon^{10})$ and therefore we have
  \[\inf{\pi\hichi}\le (z\cdot C_1+O(1))\inf{\pi\ichi}
  \le C_1\inf{\pi\ichi}\] for sufficiently large $C_1$.

  Similarly, the maximum boundary cost of $\hat\chi$ is at most
  $C_1\cdot (z\inf{\bnd\ichi}+O(\inf{\pi\ichi}^{1/p})+ C_2\cdot
  (z^{1/p}\cdot \inf{\pi\ichi})+O(\inf{\bnd\ichi}+
  \inf{\pi\ichi}^{1/p})$.  It holds
  \begin{align*}
    \inf{\bnd\hichi}
    &\le (z\cdot C_1+O(1))\inf{\bnd\ichi}\\
    &\quad+(z^{1/p}\cdot C_2+O(C_1+1))\inf{\pi\ichi}^{1/p}\\
    &\le C_1\inf{\bnd\ichi}+(z^{1/p} C_2 +O(C_1))\inf{\pi\ichi}^{1/p}
  \end{align*}
  for sufficiently large $C_1$.  Then if $C_2$ is large enough
  relative to $C_1$, the maximum boundary cost of $\hat\chi$ is at
  most $ C_1\cdot\inf{\bnd\ichi}+C_2\cdot\inf{\pi\ichi}^{1/p}$.

  The claimed running time $O(t(|G|))$ follows from the facts that $t$
  is a linear function and that the size of the considered graph
  decreases by a constant factor with each application of the
  shrinking procedure.
\end{proof}

\section{Shrinking procedure}\label{sec:shrink}
In this section we show Lemma~\ref{shrinkproc}.  Let $\epsilon>0$ be a
sufficiently small absolute constant.  The precise value of $\epsilon$
is not important. For convenience, we write $M:=\epsilon^5$, $\Psi:=
w$, $\Phi\ho 1:=\pi$, and $\Phi\ho 2:=\deg_W$, where $\deg_W(v)$ is
the degree of $v$ in $G[W]$.  Notice that $\Phi\ho 2(W_1)\le
(1-\epsilon^{10})\Phi\ho 2(W)$ implies $|G[W_1]|\le
(1-\epsilon^{10})|G[W]|$ for all vertex sets $W_1\subseteq W$.

In the following we assume $\inf \Psi\le \epsilon^{5}\Psi^*$, where
$\Psi^*:=\Psi(W)/k$ is the average weight of a color class in
coloring~$\chi:W\ra [k]$.  We need the corollaries below for our proof
of Lemma~\ref{shrinkproc}. (For the proof and lemma of the corollaries
we refer to Appendix~A.1.) %

A vertex set $U$ with $\Psi$-weight $\Theta(M\cdot\Psi^*)$ can be
partitioned into about $\Theta(M/\epsilon)$ parts, each of
$\Psi$-weight $\Theta(\epsilon\Psi^*)$.  An averaging argument shows
that for one of these parts, $X$ say, all three $\Phi\ho 1(X)$,
$\Phi\ho 2(X)$, and $\bnd(X)$ are small -- at most an
$O(\epsilon/M)$-fraction:
\begin{corollary}\label{cor1}
  For every $U\subseteq V$ with $M/2\le \Psi(U)/\Psi^*\le M$, there
  exists a subset $X$ of $U$ with $\bnd_UX=O_M(\pi^{1/p}(U))$ and
  $\epsilon \le \Psi(X)/\Psi^*\le 3\epsilon$ such that
  \begin{align*}
    \Phi\ho j(X)&\le (18\epsilon/M)\cdot \Phi\ho j(U)\\
    \bnd X&\le (18\epsilon/M)\cdot \bnd U+O_M(\pi^{1/p}(U))
  \end{align*}
\end{corollary}
Analogous to the corollary above:
\begin{corollary}\label{cor2}
  For every $U\subseteq V$ with $1/2\le \Psi(U)/\Psi^*\le M$, there
  exists a subset $X$ of $U$ with $\bnd_UX=O_M(\pi^{1/p}(U))$ and
  $\epsilon \le \Psi(X)/\Psi^*\le 3\epsilon$ such that
  \begin{align*}
    \Phi\ho j(X)&\le 18\epsilon\cdot \Phi\ho j(U),\\
    \bnd X&\le 18\epsilon\cdot \bnd U+O_M(\pi^{1/p}(U))
  \end{align*}
\end{corollary}
Somehow dual to the preceding corollaries. A vertex set $U$ is
partitioned into at most $9\Psi(U)/(\epsilon\cdot \Psi^*)$ parts, each
of $\Psi$-weight about $\epsilon/9\cdot \Psi^*$.  Among these parts,
let $X_1,X_2,X_3$ be the parts with maximum $\Phi\ho 1$-, $\Phi\ho 2$,
and $\bnd$-weight, respectively.  Then for the union $X=X_1\cup
X_2\cup X_3$, all three $\Phi\ho 1(X)$, $\Phi\ho 2(X)$ and $\bnd (X)$
are large -- at least a $(\epsilon/9\cdot \Psi^*/\Psi(U))$-fraction.
\begin{corollary}\label{cor3}
  For every $U\subseteq V$ with $\epsilon\le \Psi(U)/\Psi^*\le M$,
  there exists a subset $X$ of $U$ with $\bnd_UX=O_M(\pi^{1/p}(U))$
  and $\epsilon \le \Psi(X)/\Psi^*\le \epsilon+\inf\Psi/\Psi^*$ such
  that
  \begin{align*}
    \Phi\ho j(U\backslash X)
    &\le (1-\epsilon/9\cdot \tfrac{\Psi^*}{\Psi(U)})\cdot \Phi\ho j(U), \\
    \bnd (U\backslash X) &\le (1-\epsilon/9\cdot
    \tfrac{\Psi^*}{\Psi(U)})\cdot \bnd U+O_M(\pi^{1/p}(U))
  \end{align*}
\end{corollary}
Now we are armed to show Lemma~\ref{shrinkproc}.  We remark that sets
$X$ as in the corollaries above can be obtained in
time~$O_M(t(|G[U]|))$.
\begin{proof}[Proof of Lemma~\ref{shrinkproc}]
  We are given a coloring $\chi$ of a vertex set $W\subseteq V$ with
  $\inf{\Psi\ichi}\le M\Psi^*$
  Our aim is to find two $k$-coloring $\chi_0$ and $\chi_1$ with each
  vertex of $W$ being colored in exactly one of two colorings, such
  that $\chi_0$ is almost strictly $\Psi$-balanced and $\chi_1$ is
  weakly balanced with $\inf{\Psi\ichi_1}\le M\cdot
  \avg{\Psi\ichi_1}$.

  First, we transform coloring $\chi$ into a coloring $\tilde\chi$
  with maximum $\Psi$-weight at most $M/2\cdot \Psi^*$ and minimum
  $\Psi$-weight at least $\epsilon\cdot\Psi^*$. This transformation is
  done by moving parts generated by Corollary~\ref{cor1} and
  \ref{cor2} from ``heavy'' color classes to ``light'' color classes.
  Then Corollary~\ref{cor3} can be applied to each color class
  $\tichi(i)$. The corollary yields sets $X_i\subseteq \tichi(i)$ with
  $\Psi(X_i)$ at least $\epsilon\Psi^*$ and at most this value plus
  $\inf\Psi$.  Hence, the restriction of $\tilde\chi$ to the union
  $W_0$ of the sets $X_1$ to $X_k$ is an almost strictly balanced
  coloring.  So we can define $\chi_0:=\tilde\chi\res{W_0}$.  On the
  other hand, the restriction of $\tilde\chi$ to the complement
  $W_1:=W\setminus W_0$ is a coloring with maximum $\Psi$-weight at
  most $M/2\cdot \Psi^*$. It is not difficult to check that $M/2\cdot
  \Psi^*\le M\cdot \Psi(W_1)/k$ (cf. Claim~\ref{clm:12}).  So coloring
  $\chi_1:=\tilde\chi\res{W_1}$ fulfills $\inf{\Psi\tichi_1}\le M\cdot
  \avg{\Psi\ichi_1}$.

  We need to ensure that the construction of $\tilde\chi$ does not
  increase the $\Phi\ho j$-weight of color classes by more than a
  small fraction of $\inf{\Phi\ho j\ichi}$.  Otherwise,
  Corollary~\ref{cor3} could not guarantee $\inf{\Phi\ho j\ichi_1}\le
  (1-\epsilon^{10})\inf{\Phi\ho j\ichi}$ as required by the lemma.
  Similarly, the construction should not increase the boundary costs
  of color classes by too much.

  The idea for constructing $\tilde\chi$ is as follows.  We start with
  $\tilde\chi=\chi$.  First the maximum $\Psi$-weight of $\tilde\chi$
  is reduced to $M/2\cdot \Psi^*$.  For color classes $\tichi(i)$ with
  $\Psi$-weight larger than $M/2\cdot \Psi^*$, we uncolor subsets
  $X\subseteq \tichi(i)$ as in Corollary~\ref{cor1} and store these
  sets in a data structure called $\Buffer\subseteq 2^W$.  The
  corollary ensures that all parts $X\in \Buffer$ have $\Psi$-weight
  about $\epsilon\Psi^*$ but only very small $\Phi\ho j$-weight.  Then
  the minimum $\Psi$-weight of $\tilde\chi$ is increased to $\epsilon
  \Psi^*$.  If $\Buffer$ had enough elements, we could simply assign
  one part $X\in \Buffer$ to each color class with $\Psi$-weight less
  than $\epsilon \Psi^*$.  Otherwise, we have to use
  Corollary~\ref{cor2} to generate more parts (from color classes with
  $\Psi$-weight at least $\Psi^*/2$). Note that the parts generated by
  this corollary are $M$ times more costly than the parts generated by
  Corollary~\ref{cor1}. In case that $\Buffer$ contained more elements
  than there were color classes with weight below $\epsilon\Psi^*$, we
  distribute the remaining parts of $\Buffer$ greedily among the color
  classes.  An important observation about our construction that if we
  assigned more than one part $X$ to a color class, then all theses
  parts are as in Corollary~\ref{cor1}.

  It remains to give a detailed description of the shrinking
  procedure.  As indicated before, we start with $\tilde\chi=\chi$.
  For convenience, we subdivide the procedure into three phases
  (subroutines) \Del, \Add and \Empty.

  The procedure \Del is used to reduce the $\Psi$-weight of a color
  class by a constant fraction of $\Psi^*$.  The algorithm will
  iterate this procedure until each color class of $\tilde\chi$ has
  weight at most $M/2\cdot\Psi^*$.
  \begin{procedure}{CutDown}{color $i\in[k]$}
    \irem{Precondition: $M/2\cdot \Psi^*<\Psi\,\tichi(i)\le
      M\cdot\Psi^*$}
  \item Compute a subset $X$ of $\tichi(i)$ with $\epsilon\le
    \Psi(X)/\Psi^*\le3\epsilon$ as in Corollary~\ref{cor1}
  \item Uncolor all vertices of $X$ in $\tilde\chi$
  \item Insert set $X$ into $\Buffer$%
  \end{procedure}
  The procedure \Add increases the $\Psi$-weight of color class by
  assigning a part $X\subseteq W$ to it. Either part $X$ is an element
  of $\Buffer$, or $X$ is a subset of some color class $\tichi(j)$ as
  in Corollary~\ref{cor2}.  When procedure \Add is iterated
  appropriately, we yield a coloring $\tilde\chi$ with each color
  class having $\Psi$-weight at least $\epsilon\Psi^*$.
  \begin{procedure}{AddTo}{color $j\in[k]$}
    \irem{Precondition: $\Psi\,\tichi(j)<\epsilon\cdot\Psi^*$}
  \item If $\Buffer=\emptyset$, \subitem then let $i$ be a color with
    $\Psi\,\tichi(i)\ge \Psi^*/2$, \subitem compute a subset $X$ of
    $\tichi(i)$ with \subitem $\epsilon\le \Psi(X)/\Psi^*\le
    3\epsilon$ as in Corollary~\ref{cor2}, \newline else \subitem let
    $X$ be an arbitrary element of $\Buffer$, \subitem and remove $X$
    from $\Buffer$%
  \item Paint all vertices in $X$ with color $j$.
  \end{procedure}
  Finally, the procedure \Empty is used to empty the buffer in case
  that $\Buffer$ contained more elements than there were color classes
  with weight below $\epsilon\Psi^*$.  A part $X\in \Buffer$ is simply
  assigned to a color class with at most average $\Psi$-weight.
  \begin{procedure}{ReduceBuffer}{}
    \irem{Precondition: $\Buffer\neq \emptyset$}
  \item Remove some part $X$ from the $\Buffer$.
  \item Let $j\in [k]$ be a color with $\Psi\,\tichi(j)\le \Psi^*$
  \item Paint all vertices in $X$ with color $j$.
  \end{procedure}
  Now our shrinking procedure reads as follows.
  \begin{procedure}{Shrink}{coloring $\chi\col W\ra [k]$}
    \irem{Precondition: $\inf{\Psi\ichi}\le M\Psi^*$}
  \item Start with $\tilde\chi\la \chi$, and $\Buffer\la
    \emptyset\subseteq 2^W$.
  \item As long as %
    $\exists$
    a color $i$ with $\Psi\,\tichi(i)>M/2\cdot\Psi^*$,
    \subitem do \Del($i$).
  \item For every color $i$ with $\Psi\,\tichi(i)<\epsilon\Psi^*$,
    \subitem do \Add($i$).
  \item Until $\Buffer= \emptyset$, \subitem do \Empty().
    \irem{Assert: $\tilde\chi$ is a total coloring of $W$ with
      $\epsilon \Psi^*\le \Psi\tichi(i)\le M/2\cdot \Psi^*$ for all
      $i\in[k]$.}
  \item For each color $i\in[k]$, \subitem compute a subset $X_i$ of
    $\tichi(i)$ as in Cor.~\ref{cor3}\subitem with
    $\epsilon\Psi^*\le\Psi(X_i)\le \epsilon\Psi^*+\inf\Psi$.
  \item Set $W_0:=X_1\cup\ldots\cup X_k$ and $W_1:=W\setminus W_0$.
  \item Return the colorings $\chi_0:=\tilde\chi\res{W_0}$ and
    $\chi_1:=\tilde\chi\res{W_1}$.
  \end{procedure}
  The assertion before step~$(5.)$, which is easily seen to hold, and
  the fact $\Psi(X_i)-\epsilon\Psi^*\in[0,\inf \Psi] \subseteq
  [0,\epsilon^5\Psi^*]$ imply the claim below (for sufficiently small
  $\epsilon$).
  \begin{claim}\label{clm:12}
    If $\chi_0$ and $\chi_1$ are the colorings computed by
    \Shrink($\chi$), then $\chi_0$ is almost strictly balanced with
    $\epsilon\Psi^*\le \Psi\ichi_0(i) =\Psi(X_i)\le
    \epsilon\Psi^*+\inf \Psi$ and $\chi_1$ satisfies
    $\inf{\Psi\ichi_1}\le M\cdot \Psi_1^*$, where
    $\Psi_1^*:=\Psi(W_1)/k$ is the average $\Psi$-weight of a
    $k$-coloring of $W_1$.
  \end{claim}
  \begin{pf}
    By construction, coloring $\chi_0$ fulfills the claim.  To see
    that the claim holds for $\chi_1$, we observe $\Psi(W_1)/k\ge
    \Psi(W)/k-\epsilon\Psi^*-\inf \Psi$.  So it holds $\Psi^*_1\ge
    (1-\epsilon-\epsilon^5)\Psi^*$, and therefore
    $\inf{\Psi\ichi_1}\le M/2\cdot\Psi^*\le M\Psi^*_1$ for all
    sufficiently small $\epsilon>0$.  \smallqed
  \end{pf}

  The next claim observes a simple but crucial property of the
  algorithm.  The colors are naturally divided into donators and
  receivers.
  Formally, let $\Source\subseteq [k]$ be the set of colors $i$ with
  $\ichi(i)\setminus \tichi(i)\neq \emptyset$, and $\Sink\subseteq
  [k]$ be the set of colors $j$ with $\tichi(j)\setminus \ichi(j)\neq
  \emptyset$.

  Clearly, $\Source$ consists exactly of the colors for which we
  called procedure $\Del$, or that were selected in step $(1.)$ of
  $\Add$.  Similarly, $\Sink$ consists of all colors for which we
  called procedure $\Add$, or that were selected in step $(2.)$ of
  $\Empty$.
  For sufficiently small $\epsilon$, no color can be both a source and
  a sink:
  \begin{claim}\label{clm:11}
    $\Source\cap \Sink=\emptyset$
  \end{claim}
  \begin{pf}
    We distinguish two cases.

    First, we consider the case that $\Del$ created more
    $\Buffer$-parts than were used by $\Add$ -- so $\Empty$ was called
    at least once. Then $\Source$ consists only of those colors $i\in
    [k]$ for which $\Del(i)$ has been called. Hence for all colors
    $i\in \Source$, \[\Psi\tichi(i)\ge (M/2-3\epsilon)\Psi^*>\Psi^*\]
    is an invariant of the algorithm.  From this invariant it follows
    that $\Source$ contains no color $j\in[k]$ for which $\Add(j)$ was
    called or that was selected in step $(2)$ of procedure \Empty. So,
    $\Source$ and $\Sink$ are disjoint sets.

    In the case that $\Empty$ was never called by the algorithm,
    $\Sink$ consists only of those colors $j\in[k]$ for which
    $\Add(j)$ was called. So for all colors $j\in \Sink$,
    \[
    \Psi\tichi(j)\le (\epsilon+3\epsilon)\Psi^*<\Psi^*/2
    \]
    is an invariant of the algorithm. This invariant implies that
    $\Source$ cannot contain a color $i\in[k]$ for which $\Del(i)$ was
    called or that was selected in step $(1)$ of procedure
    $\Add$. Therefore, $\Source$ and $\Sink$ must be mutually
    exclusive.  \smallqed
  \end{pf}
  Based on Claim~\ref{clm:11}, we show bounds on the $\Phi\ho
  j$-weight and boundary cost of the color classes in $\tilde\chi$.
  These bounds will imply that the maximum $\Phi\ho j$-measure and the
  maximum boundary cost of both $\chi_0$ and $\chi_1$ are as required
  by the lemma.

  Consider any source color $i\in \Source$.  Every time when its color
  class is changed by the algorithm, the $\Psi$-weight of $\tichi(i)$
  decreases by at least $\epsilon\Psi^*$.  Since the initial weight is
  at most $M\cdot \Psi^*$, there can be at most $M/\epsilon$ such
  changes.  Thus, the increase of $\bnd\tichi(i)$ is bounded by
  $M/\epsilon\cdot
  O_M(\pi^{1/p}(\ichi(i)))=O_M(\inf{\pi\ichi}^{1/p})$, because each
  applied cut has cost $O_M(\pi^{1/p}(\tichi(i)))$.  This observation
  shall imply that for the analysis of our shrinking algorithm, the
  boundary cost function $\bnd$ behaves like one of the vertex
  measures $\Phi\ho j$ modulo additive terms of order
  $\inf{\pi\ichi}^{1/p}$.  So we restrict ourselves in the following
  to show the requirements of the lemma only for $\Phi\ho j$. The
  arguments for $\bnd$ are analogous.

  Consider a non-sink~color~$i$, and let $U=\tichi(i)$ be its color
  class in step~$(5.)$ of procedure~\Shrink.  By the assertion before
  step~$(5.)$, it holds $\Psi^*/\Psi U\ge 2/M$.  Thus
  $\ichi_1(i)=U\setminus X_i$ has $\Phi\ho j$-weight at most
  $(1-\frac{2\epsilon}{9M})\Phi\ho j(U)\le (1-\epsilon^{-10})\Phi\ho j
  (U)$ by Corollary~\ref{cor3}.  Since the considered color~$i$ is not
  a $\Sink$, we have $\Phi\ho j\tichi(i)\le \inf{\Phi\ho j\ichi}$
  throughout the algorithm. So $\Phi\ho j\ichi_1(i)$ is as required
  for $i\not\in \Sink$.
  \begin{claim}
    For non-sink colors~$i$ and all $j\in[r]$, it holds $\Phi\ho
    j\ichi_0(i)\le \inf{\Phi\ho j\ichi}$ and $\Phi\ho j\ichi_1(i)\le
    (1-\epsilon^{10})\inf{\Phi\ho j\ichi}$
  \end{claim}
  To show the corresponding claim for sink colors, we need to estimate
  the $\Phi\ho j$-weight of the parts $X\subseteq W$ that get
  transfered from $\Source$ colors to $\Sink$ colors.

  Two types of parts are considered by the algorithm.  By
  Corollary~\ref{cor1}, we have for any part $X$ that gets inserted
  into $\Buffer$,
  \begin{equation}
    \label{eq:6}
    \Phi\ho j(X)\le 18\epsilon/M\cdot \inf{\Phi\ho j\ichi},  
  \end{equation}
  and by Corollary~\ref{cor2}, any parts $X$ that is painted in
  step~$(2.)$ of procedure $\Add$ satisfies
  \begin{equation}
    \label{eq:7}
    \Phi\ho j(X)\le 18\epsilon\cdot \inf{\Phi\ho j\ichi}.
  \end{equation}

  Since the parts of the second type are much more expensive (in terms
  of $\Phi\ho j$-weight) than the parts of the first type only the
  following observations allows us to derive the required bound on
  $\Phi\ho j\ichi_1(i)$ for $i\in Sink$.

  \emph{Key-Observation:} For a sink color $i$, either a) all received
  parts are of the first type, or b) throughout the algorithm color
  $i$ received only one part and hence $\Psi\tichi(i)\le \epsilon
  \Psi^*+3\epsilon\Psi^*$, because
  this parts had $\Psi$-weight at most $3\epsilon\Psi^*$.

  We show the required bound on $\Phi\ho j\ichi_1(i)$ by
  distinguishing these two cases.

  \emph{Case a):} It is an invariant of the algorithm that sink colors
  have color classes of $\Psi$-weight at most
  $\Psi^*+3\epsilon\Psi^*\le 2\Psi^*$.  And since each received part
  has $\Psi$-weight at least $\epsilon\Psi$, a sink color~$i$ can
  receive no more than $2/\epsilon$ parts.  In the current case, all
  of these parts have $\Phi\ho j$-weight at most
  $18\epsilon/M\cdot\inf{\Phi\ho j\ichi}$.  Hence in step~$(5.)$ of
  procedure~\Shrink, it holds $\Phi\ho j(U)\le (1+36/M)\inf{\Phi\ho
    j\ichi}$ for the $i$-th color class $U=\tichi(i)$.  Since
  $\Psi^*/\Psi(U)\ge 1/2$, we have by Corollary~\ref{cor3}
  \[
  \Phi\ho j(U\backslash X_i)/\inf{\Phi\ho j\ichi}\le
  (1-\epsilon/18)(1+36\epsilon^5)\le 1-\epsilon^{10}
  \]
  for sufficiently small $\epsilon>0$. Thus the $\Phi\ho j$-weight of
  $\ichi_1(i)=U\setminus X_i$ is as required by the lemma.

  \emph{Case b):} By our key observation, we have $\Psi^*/\Psi(U)\ge
  1/4\epsilon$ for the $i$-th color class $U=\tichi(i)$ at the end of
  procedure \Shrink. It also holds $\Phi\ho j(U)\le
  (1+18\epsilon)\inf{\Phi\ho j\ichi}$, since only one part was
  received by color $i$ in the course of the algorithm.  By
  Corollary~\ref{cor3}, the $\Phi\ho j$-weight of
  $\ichi_1(i)=U\setminus X_i$ satisfies for sufficiently small
  $\epsilon>0$, \[ \Phi\ho j(U\backslash X_i)/\inf{\Phi\ho j\ichi}\le
  (1-\tfrac{\epsilon}{9}\cdot \tfrac{1}{4\epsilon})(1+18\epsilon)\le
  1-\epsilon^{10}
  \]

  This case distinction showed for every $i\in Sink$:
  \begin{claim}
    $\Phi\ho j\ichi_1(i)\le (1-\epsilon^{10})\inf{\Phi\ho j\ichi}$.
  \end{claim}

  The discussion for case a) above, implies that any sink color $i$
  receives at most constantly many parts. Then it follows from
  relations~\eqref{eq:6} and \eqref{eq:7}, that $\Phi\ho
  j\ichi_1(i)=O_M(\inf{\Phi\ho j\ichi})$.

  The fact that any color class is altered only constantly often also
  shows that the algorithm can be implemented to run in time
  $O_M(t(|G[W]|))$, provided that appropriate data structures are
  used.  For example in step~$(2.)$ of procedure~\Shrink, the colors
  with $\Psi\tichi(i)>M/2\cdot \Psi^*$ should be maintained by a
  stack, so that such colors can be found in constant time.
\end{proof}

\section{Splittability of grid graphs}\label{sec:splitt-grid-graphs}

For the case of unit-costs, many graph classes like fixed-minor free
graphs and finite element meshes have bounded $p$-splittability for
some $p>1$ (cf. Appendix~\ref{lowersec}, Remark~\ref{resultlist}).  In
case of arbitrary edge costs, only planar graphs were known to have
splitting sets of low costs.

There is a naive way to generalize existing splittability results for
the unit costs case to the case of arbitrary edge costs.  Obviously,
it holds $\sigma_p(G,c)\le \sigma_p(G,\bbbone_E)\cdot \inf c \cdot
\inf{1/c}$ for every graph $G=(V,E)$ and arbitrary edge costs $c\col
E\ra \Rp$ (assuming without loss of generality $c(e)>0$ for all edges
$e\in E)$.  In this section, we show that the situation for
$d$-dimensional ``grid graphs'' is better.  A \emph{grid graph} in a
$d$-dimensional space is a graph $G=(V,E)$ with $V\subseteq \Z^d$ and
$\onorm{\vec x-\vec y}=1$ for all edges $\{\vec x,\vec y\}\in E$.  The
theorem of this section implies that
$\sigma_p(G,c)=O_d(\log^{1/d}(\phi)\cdot \sigma_p(G,\bbbone))$ if $G$
is a $d$-dimensional grid, $p=d/(d-1)$ and $\phi:=\inf c\cdot \inf
{1/c}$ is the ratio of the maximum cost of an edge to the minimum cost
of an edge.

Although grids form a very restrictive graph family, many graphs
arising in practical applications, e.g., in climate simulation
(cf. Introduction), are ``close'' to grid graphs and might be embedded
into grids such that boundary costs are preserved up to constant
factors.

Furthermore, the results in this section can be seen as a starting
point for further investigations of the splittability for more general
non-planar graphs with arbitrary edge costs.  Note that for $d\ge 3$,
the class of $d$-dimensional grids does not exclude any minor and
hence is ``far'' from being planar.

We state the main theorem of this section.
\begin{theorem}\label{mainthm}
  Let $G=(V,E)$ be a $d$-dimensional grid graph with edge costs
  $c:E\ra \Rp$ and vertex weights $w:V\ra \Rnn$.  Then, for all
  splitting values $w^*$ with $0\le w^*\le \onorm w$, there exists a
  $w^*$-splitting set $U\subseteq V$ of cost
  \[O(d\cdot\log^{1/d}(\phi+1)\cdot \pnorm c)\] where $p:=d/(d-1)$ and
  $\phi:= \max_E c/\min_E c$ is the \emph{fluctuation} of $c$.  Such a
  splitting set can be computed in time $O(n\log \phi)$.

  And since the class of $d$-dimensional grid graphs is closed under
  taking subgraphs, we have \[\sigma_p(G,c)=O_d(\log^{1/d}(\phi+1)).\]
\end{theorem}

\newcommand{\vp}{\varphi}

In the following, let $G=(V,E)$, $c\col E\ra \Rp$, $w\col V\ra \Rnn$,
and $w^*$ be as in Theorem~\ref{mainthm}.  We aim for a recursive
algorithm to find a $w^*$-splitting set in $G$ with small boundary
cost.

The idea of the algorithm is as follows.  We consider a \emph{coarser}
graph that is obtained by identifying vertices of $G$, i.e., for a
mapping $\vp\col V\ra \Z^d$, the coarser graph $G/\vp=(V/\vp,E/\vp)$
has the non-empty pre-images $\vp^{-1}(\vec a)=\{\vec x\in V\mid
\vp(\vec x)=\vec a\}\in V/\vp$ as nodes and contains an arc
$\{\vp^{-1}(\vec a),\vp^{-1}(\vec b)\}\in E/\vp$ if and only if an
edge $\{\vec x,\vec y\}$ of $G$ connects the two disjoint sets
$\vp^{-1}(\vec a)$ and $\vp^{-1}(\vec b)$.  Formally,
\[
V/\vp := \Big\{\vp^{-1}(\vec a)\neq \emptyset \mid \vec a\in
\Z^d\Big\},
\]
\[
E/\vp:=\Big\{ \{Q ,R\} \subseteq V/\vp \mid \vec x\in Q, \vec y\in
R,\{\vec x,\vec y\}\in E, Q\neq R\Big\}
\]
The weights and costs are translated to $G/\vp$ in a straight-forward
manner; we define $w/\vp(Q):=w(Q)$ for each $Q\in V/\vp$ and
$c/\vp(Q,R):=\sum_{\vec x\in Q,\vec y\in R}c(\vec x,\vec y)$.  In this
coarser graph, which shall also be a grid, we then use a trivial
algorithm to find a splitting set $\mathcal S\subseteq V/\vp$ and a
node $Q\in V/\vp\setminus \mathcal S$ with $w/\phi (\mathcal S)\le
w^*< w/\phi(\mathcal S)+w(Q)$.  So $\mathcal S$ has the desired weight
$w^*$ up to the weight of $Q$.  Now the idea is to proceed recursively
in order to compute a $(w^*-w/\vp(\mathcal S))$-splitting set
$U'\subseteq Q$ in $G':=G[Q]$.  Then, $U:=\bigcup \mathcal S\cup U'$
will be the required $w^*$-splitting set in $G$.

To bound the boundary cost $\bnd U$ of the resulting splitting set,
our reasoning will be as follows.  Since any boundary edge $\delta(U)$
is contained in either $\delta(\bigcup \mathcal S)$, $\delta(Q)$, or
$\delta_{G[Q]}(U')$, the boundary cost satisfies $\bnd U\le \bnd
(\bigcup \mathcal S\cup Q)+\bnd_{Q} U'$. In our analysis we shall use
the crude over-estimate $\bnd(\bigcup\mathcal S\cup Q)\le \onorm
{c/\vp}$ to deduce that $\bnd U$ is at most $\onorm{c/\vp}+\bnd_Q U'$.

As we will see, there are canonical choices of $\vp$ that guarantee a
good upper bound on $\onorm{c/\vp}$ (cf. Lemma~\ref{lemcheap}).
However, to make the recursion work, we need to ensure that a
splitting set in $G'=G[Q]$ is somehow easier to obtain and less
costly.  It seems difficult to do so, and therefore we circumvent this
issue.  The recursive instance does not use the same edge costs $c$
but ``reduced'' edge costs $c'\col E'\ra \Rp$ with $c'_e:=(c_e-1)/2$
for all edges with $c_e>1$.  All edges with $c_e\le 1$ are discarded
in $G'$.  With this choice of $c'$, we end up with an empty graph
after $O(\log \inf c)$ levels of recursion.  So there is some notable
progress when going to recursive instances.  On the other hand, we
need to take into account the edges of $G$ that were discarded in
$G'$, when we want to deduce a bound on $\bnd_Q U'$ from $\bnd'(U')$,
where $\bnd'$ denotes the boundary costs in $G'$ with respect to $c'$.
We use again a rough estimate $\bnd_Q U'\le |\delta_{G[Q]}(U')|+2
\bnd' U'$, which holds with equality only if $\delta_{G[Q]}(U')$
contains no edge $e$ with $c(e)<1$.  In order to control
$|\delta_{G[Q]}(U')|$, we shall use specific properties of $Q\in
V/\vp$ that follow from our choice of $G/\vp$, and we will exploit an
invariant of our recursive algorithm, namely that the computed
splitting sets are ``monotone''
(cf. Lemma~\ref{lemsmall}-\ref{lemmon}).

\paragraph*{Obtaining a coarser graph with low edge costs.}
Grid graphs can be coarsened nicely in a geometrically intuitive way.
We partition the $d$-dimensional space $\R^d$ in ``half-open''
hyper-cubes $\vec x+[0,\,\ell)^d$ of measure $\ell^d$, where $\vec
x+Y:=\{\vec x+\vec y\mid\vec y\in Y\}$ denotes the Minkowski sum of
$\vec x\in \R^d$ and $Y\subseteq \R^d$.  This partition is in such a
way that each face of a cube in the partition is also the face of
another cube.  Then, we identify all vertices of grid $G$ that lie in
the same cube.

Formally, we define mappings \[ \vp\ho \ell_\alpha\col \Z^d\ra
\Z^d,\quad \vec a\mapsto\floor{(\vec a+(\alpha-1)\cdot
  \bbbone_d)/\ell}\] for positive integers $\alpha,\ell\in \N$, where
$\bbbone_d\in \R^d$ is the vector of all ones and $\floor{\vec x}$ is
the integer vector with components $\floor{x_i}$. Then, the graph
$G/\vp\ho \ell_\alpha$ is a coarser graph of the kind described above.

We say that a graph $G/\vp$ is \emph{$\ell$-coarse} if $\vp=\vp\ho
\ell_\alpha$ for some $\alpha\in [\ell]$.  The lemma below observes
that we can always find an $\ell$-coarse graph with low edge costs.
\begin{lemma}\label{lemcheap}
  For each positive integer $\ell$, there exists an $\ell$-coarse
  graph $G/\vp$ with \[ \onorm{c/\vp}\le \onorm{c}/\ell\]
\end{lemma}
\begin{proof}
  Since each edge of $G$ accounts for exactly one of the cost
  functions $c/\vp\ho \ell_1$ through $c/\vp\ho \ell_\ell$, we have
  $\onorm{c/\vp\ho \ell_1}+\ldots+\onorm{c/\vp\ho
    \ell_\ell}=\onorm{c}$. Hence, $\onorm{c/\vp\ho \ell_\alpha}\le
  \onorm{c}/\ell$ for some $\alpha\in[\ell]$.
\end{proof}

We say that an $\ell$-coarse graph $G/\vp$ is \emph{cheap} if
$\onorm{c/\vp}\le \onorm{c}/\ell$.

\paragraph*{Monotone sets.} For a set $Q\subseteq \Z^d$, we say that a
subset $W\subseteq Q$ is \emph{monotone} (in $Q$), if for all $\vec
x\in Q$ and $\vec y\in W$ with $\vec x\le \vec y$, it holds $\vec x\in
Q$, where $\vec x\le \vec y$ means that we have $x_i\le y_i$ for each
component of $\vec x$ and $\vec y$.

We shall see later that it is an invariant of our recursive algorithm
that the computed splitting set $U'$ of $G'=G[Q]$ is monotone in $Q\in
V/\vp$.  The next lemma allows us to bound $|\delta_{G[Q]}(U')|$,
provided that $G/\vp$ is $\ell$-coarse.

\begin{lemma}\label{lemsmall}
  Let $Q\subseteq \Z^d$ be a node of an $\ell$-coarse graph $G/\vp$,
  i.e., $Q\subseteq \vec x+[0,\ell)^d$ for some $\vec x\in \Z^d$.

  Then for any monotone set $W$ in $Q$, it holds
  $|\delta_{G[Q]}(W)|\le d\ell^{d-1}$.
\end{lemma}
\begin{proof}
  For each edge $\vec a \vec b\in \delta_{G[Q]}(W)$, consider the line
  $L=\{\lambda \vec a+(1-\lambda)\vec b \mid \lambda\in \R\}$ through
  this edge.  By monotonicity of $W$, no other edge of
  $\delta_{G[Q]}(W)$ is contained in $L$. On the other hand, we can
  uniquely identify $L$ by the point in $Q\cap L$ with the smallest
  coordinates (the point at which $L$ ``leaves'' $Q$).  Notice that
  the points in $L\cap Q$ can differ only in one component and
  therefore agree in $d-1$ components. So if the direction of $L$ is
  fixed, the ``lowest'' point in $L\cap Q$ is uniquely determined by
  its coordinates in $d-1$ components.

  Any of these ``leaving'' points can be generated as follows.  First,
  we select one of the $d$ components, say the $i$-th component, that
  needs to be smallest.  Then, there are at most $\ell^{d-1}$ choices
  for the entries of the other components.  No further choices remain,
  since the $i$-th component needs to be smallest.  Thus, there are at
  most $d\cdot \ell^{d-1}$ edges in the cut $\delta_{G[Q]}(W)$.
\end{proof}

The lemmata below shall imply the invariant that $U'$
is a monotone set in $Q$.

\begin{lemma}\label{lemlex}
  If $\vec a_1,\ldots,\vec a_n$ is a lexicographic ordering of a set
  $S\subseteq \Z^d$, then for all $i\in[n]$, the subset $\{\vec
  a_1,\ldots,\vec a_i\}$ is monotone in $S$.
\end{lemma}
\begin{proof}
  A vector $\vec x$ is lexicographically less than a vector $\vec y$
  if it holds $\vec x\le \vec y$.
\end{proof}

For every node $R=\vp^{-1}(\vec a)\subseteq \Z^d$ of $G/\vp$, we
conveniently define $\vp(R):=\vec a$.

\begin{lemma}\label{lemmon}
  Let $G/\vp$ be an $\ell$-coarse graph of grid $G=(V,E)$ and
  $Q_1,\ldots,Q_i\in V/\vp$ be a sequence of distinct nodes of
  $G/\vp$. Suppose both $\{\vp(Q_1),\ldots,\vp(Q_{i-1})\}$ and
  $\{\vp(Q_1),\ldots,\vp(Q_{i})\}$ are monotone sets in
  $\vp(V/\vp):=\{\vp(R)\mid R\in V/\vp\}$.  Then for any monotone set
  $W$ in $Q_i$, the set $Q_1\cup \ldots\cup Q_{i-1}\cup W$ is monotone
  in $V$.
\end{lemma}
\begin{proof}
  Notice that $Q_1\cup \ldots \cup Q_{i-1}$ is monotone in $V$, since
  $\{\vp(Q_1),\ldots,\vp(Q_{i-1})\}$ is monotone in $\vp(V/\vp)$ and
  $\vp$ is monotone, i.e., $\vp(\vec x)\le \vp(\vec y)$ for all $\vec
  x\le \vec y$.  And since $W$ is monotone in $Q_i$, it only remains
  to verify that for all $\vec y\in W$ and $\vec x\in V \setminus W$
  with $\vec x\le \vec y$, it holds $\vec x\in Q_j$ for some
  $j<i$. But this follows from the monotonicity of
  $\{\vp(Q_1),\ldots,\vp(Q_i)\}$ in $\vp(V/\vp)$ and from the
  monotonicity of $\vp$.
\end{proof}

\paragraph*{Final algorithm.}
We now have gathered all ingredients of the algorithm for computing
monotone splitting sets in grid graphs.  However, we have not yet
determined how $\ell$ should be chosen.  Remember that we derived the
relation \[ \bnd U\le \onorm{c/\vp}+|\delta_{G[Q]}(U')|+2\bnd'U'
\] between the cost of $U$ in $G$ and the cost of $U'$ in $G'$.  From
Lemma~\ref{lemcheap} and Lemma~\ref{lemsmall}, it would follow $\bnd
U\le \onorm{c}/\ell+d\ell^{d-1}+2\bnd'U'$. When we choose
$\ell:=(\onorm{c}/d)^{1/d}$, the expression is minimized and we obtain
\begin{equation}
  \bnd U\le 2d^{1/d}\onorm{c}^{1-1/d} + 2\cdot\bnd'U'.\label{eq:18}
\end{equation}
\newcommand{\GridSplit}{\textsc{GridSplit} }
\begin{procedure}{\GridSplit}{grid graph $G=(V,E)$, edge costs $c\col
    E\ra \Rp$, $w^*\in \Rnn$}
\item Compute a cheap $\ell$-coarse graph $G/\vp$ of $G$ with
  $\ell:=\max\{\lceil(\onorm{c}/d)^{1/d}\rceil,1\}\ \ $
  (Lem.~\ref{lemcheap})
\item Find an ordering $Q_1,\ldots,Q_q$ of the vertices of $G/\vp$
  such that $\vp(Q_j)$ is lexicographically less than $\vp(Q_{j+1})$
  for all $j\in[q-1]$
\item Determine a set $\mathcal S=\{Q_1,\ldots,Q_{i-1}\}$ with
  $w(\bigcup \mathcal S)\le w^*<w(\bigcup\mathcal S)+w(Q_i)$.
\item \emph{Trivial case:} If $\ell=1$ then return a $w^*$-splitting
  set among $\bigcup\mathcal S$ and $\bigcup \mathcal S\cup Q_i$
\item Recursively compute a monotone splitting set \[
  U':=\GridSplit\Big(G',c',w^*-w(\bigcup\mathcal S\Big),\] where
  $G':=G[Q]\setminus \{e\in E\mid c(e)\le 1\}$ and $c'_e:=(c_e-1)/2$
  for all $e\in E(G')$
\item Return the $w^*$-splitting set $U:=\bigcup \mathcal S\cup U'$
\end{procedure}
Notice that \GridSplit terminates after $O(\log \inf c)$ levels of
recursion. The maximum cost of an edge decreases by a factor of at
least $2$ with every level. And as soon as $\inf c\le 1$, we have
$\inf {c'}=0$ and terminate in the next level as
$\ell'=\max\{\lceil(\onorm{c'}/d)^{1/d}\rceil,1\}=1$.

Before bounding the cost of the splitting set computed by \GridSplit,
we show that the splitting set is indeed monotone in $V$.  The
monotonicity of the splitting set shall follow from Lemma~\ref{lemlex}
and Lemma~\ref{lemmon}.

\begin{lemma}\label{lemmonsplit}
  The splitting set computed by \GridSplit is monotone in $V$.
\end{lemma}
\begin{proof} By induction on $\inf c$.

  In the case $\ell=1$, the $1$-coarse graph $G/\vp$ coincides with
  $G$, since $\vp$ is the identity.  Now $\vp(Q_1),\ldots,\vp(Q_q)$ is
  a lexicographic ordering of $V$.  By Lemma~\ref{lemlex}, both
  $\bigcup \mathcal S=\{\vp(Q_1),\ldots,\vp(Q_{i-1})\}$ and
  $\bigcup\mathcal S\cup Q_i=\{\vp(Q_1),\ldots,\vp(Q_i)\}$ are
  monotone in $\vp(V/\vp)=V$.

  For $\ell>1$, the splitting set $U'$ is monotone in $Q$ by induction
  hypothesis. Also, it holds that both
  $\{\vp(Q_1),\ldots,\vp(Q_{i-1})\}$ and
  $\{\vp(Q_1),\ldots,\vp(Q_{i})\}$ are monotone in $\vp(V/\vp)$ by
  Lemma~\ref{lemlex}.  So we can apply Lemma~\ref{lemmon} and obtain
  that $\bigcup \mathcal S\cup U'=Q_1\cup\ldots\cup Q_{i-1}\cup U'$ is
  monotone in $V$.
\end{proof}

Now we are armed to show our first (technical) bound on the boundary
cost $\partial U$, which is obtained by unfolding the recurrence
\eqref{eq:18} of procedure $\GridSplit$.

\begin{lemma}
  If \GridSplit is applied to a grid graph $G=(V,E)$ with edge costs
  $c\col E\ra \Rnn$, then the returned monotone splitting set
  $U\subseteq V$ satisfies
  \[
  \bnd U\le 2^dd^{1/d}\Big(\inf c+1+\sum_{0\le i\le \log(\inf c+1)}
  2^{i/d}\onorm{c\res {E_i}}^{1-1/d}\Big)
  \]
  where $E_i:=\{e\in E\mid c(e)\ge 2^i-1\}$.
\end{lemma}
\begin{proof}
  By induction on $\lfloor\log(\inf c+1)\rfloor$.

  For $\ell=1$, we have $\onorm c\le d^{1/d}$ and therefore $\bnd U
  \le d^{1/d}$.  So, we can assume $\ell=\lceil(\onorm
  c/d)^{1/d}\rceil> 1$ and thus $\onorm{c}/d\le \ell^d\le
  2^d\onorm{c}/d$.

  In case of $\inf c\le 1$, it holds $\bnd_Q
  U'\le|\delta_{G[Q]}(U')|$.  Since $U'$ is monotone by
  Lemma~\ref{lemmonsplit} in $Q$, we have by Lemma~\ref{lemsmall} that
  $|\delta_{G[Q]}(U')|\le d\ell^{d-1}\le
  2^{d-1}d^{1/d}\onorm{c}^{1-1/d}$.  As $G/\vp$ is a cheap
  $\ell$-coarse grid, the edge costs $\onorm{c/\vp}$ are at most
  $\onorm{c}/\ell\le d^{1/d}\onorm{c}^{1-1/d}$ and thus we have $\bnd
  U\le \onorm{c/\vp}+\bnd_Q U'\le 2^{d}d^{1/d}\onorm{c}^{1-1/d}$ as
  required.

  For the rest of the proof, we can assume $\inf c>1$ and $\bnd
  U\le2^{d}d^{1/d}\onorm{c}^{1-1/d}+2\bnd' U'$ where $\bnd'$ are the
  boundary costs in $G'$ with respect to edge costs $c'$.  By
  induction hypothesis, it holds $\bnd'U'\le 2^dd^{1/d}(\inf
  {c'}+1+\sum_{0\le i\le r} 2^{i/d}\onorm{c'\res {E'_i}}^{1-1/d})$,
  where $r:= \log(\inf c'+1)=\log (\inf c+1)-1$ and $E'_i:=\{e\in
  E'\mid 2^i-1\le c'_e = (c_e-1)/2\}=E_{i+1}$. So, it holds
  $\onorm{c'\res{E'_i}}\le \onorm{c\res{E_{i+1}}}/2$ and thus $2\cdot
  2^{i/d}\onorm{c'\res {E'_i}}^{1-1/d}\le
  2^{(i+1)/d}\onorm{c\res{E_{i+1}}}^{1-1/d}$. Therefore, we have
  \[\bnd U\le 2^{d}d^{1/d}\onorm{c}^{1-1/d}+ 2^dd^{1/d}\Big(\inf
  c+1+\sum_{0\le i\le r} 2^{{i+1}/d}\onorm{c\res
    {E_{i+1}}}^{1-1/d}\Big)\] as required.
\end{proof}

Using H\"older's inequality, the lemma below arrives at a bound on
$\sum_i 2^{i/d}\onorm{c\res{E_i}}^{1-1/d}$.  Notice that by scaling
the edge cost, we can achieve $\inf{1/c}=1$ and therefore
$\phi=\inf{c}\cdot\inf{1/c}=\inf c$. So the next lemma implies the
bounds on the $p$-splittability of grid graphs from
Theorem~\ref{mainthm}.

\begin{lemma}
  For edge costs $c\col E\ra \Rp$ with $\inf{1/c}=1$, it holds
  \[
  \sum_{i=0}^{ \floor{\log (\inf c+1)}}
  2^{i/d}\onorm{c\res{E_i}}^{1-1/d}=O_d\Big( (\log^{1/d} (2\inf
  c)\cdot \norm{c}_{d/(d-1)}\Big)
  \]
\end{lemma}
\begin{proof}
  Let $r\col E\ra \N_0$ be the function that assigns each edge $e\in
  E$ to the largest index $r(e):=\floor{\log (c_e+1)}$ such that $e\in
  E_{r(e)}$. Let $s:=\inf r\le \log(2\inf c)$ denote the largest index
  with $E_s\neq \emptyset$.
  Since $c_e\ge 1$ for each edge $e\in E$, it holds $r(e)\le \log
  (2c_e)$ and thus \eq{\label{eq1} c_e\cdot \sum_{0\le i\le
      r(e)}2^{i/(d-1)}=O_d(c_e\cdot 2^{r(e)/(d-1)})=O_d(c_e^{d/(d-1)})
  } Hence, the following sum satisfies \eq{\label{eq2} \sum_{0\le i\le
      s} 2^{i/(d-1)}\onorm{c\res{E_i}}=\sum_{e\in E} c_e\sum_{0\le
      i\le r(e)} 2^{i/(d-1)}\stackrel{\eqref{eq1}}=O_d(\sum_{e\in
      E}c_e^{d/(d-1)}) } Using H\"older's inequality and
  relation~\eqref{eq2} yields for $p=d/(d-1)$ (and $q=d$),
  \eq{\label{eq3} \sum_{ i=0}^{s} 1\cdot
    (2^{i/(d-1)}\onorm{c\res{E_i}})^{1-1/d} =\inf r^{1/d}\cdot (\sum_{
      i=0}^{s}
    2^{i/(d-1)}\onorm{c\res{E_i}})^{1-1/d}\stackrel{\eqref{eq2}}=O_d(s^{1/d}
    \norm{c}_p) } From relation~\eqref{eq3}, the lemma follows as $
  2^{i/d}\onorm{c\res{E_i}}^{1-1/d}= 1\cdot
  (2^{i/(d-1)}\onorm{c\res{E_i}})^{1-1/d}$.
\end{proof}

We conclude the section with an analysis of the running time of
procedure $\GridSplit$.

\begin{lemma}
  Procedure $\GridSplit$ runs in time $O(m\log \phi)$ for a connected
  grid graph $G$ of size $m$ with edge costs $c$ of fluctuation
  $\phi$.
\end{lemma}
\begin{proof}
  We assume that the edge costs are scaled in such a way that the
  minimum edge cost is equal to $1$ and so it holds $\phi=\inf c$.
  Then the number of iterations is bounded by $O(\log \phi)$.

  It remains to show that the running time of one iteration is linear
  in the size of the grid.  Steps $(3)$-$(6)$ are easily seen to run
  in linear time. In step (1), finding a cheap $\ell$-coarse graph is
  trivial if $\ell$ is much larger than the size of $G$, because for a
  connected grid, one of the mappings $\vp\ho m_\alpha$ assigns the
  same point to all grid vertices.

  So we can assume $\ell=O(m)$ for finding a cheap $\ell$-coarse graph
  in step (1).  For each edge $\vec {ab}\subseteq \Z^d$ with
  $\onorm{\vec a-\vec b}=1$, we can determine in constant time the
  index $\alpha\in[\ell]$ with $\vp\ho \ell_\alpha(\vec a)\neq
  \vp\ho\ell_\alpha(\vec b)$.  Thus, the function $f\col [\ell]\ra
  \Rnn$ with $f(\alpha):=\onorm{c/\vp\ho \ell_\alpha}$ can be computed
  in linear time by scanning through all edges of $G$.  Now we can
  find a cheap $\ell$-coarse graph by finding the minimum of
  $f(\alpha)$ over all $\alpha\in[\ell]$.

  In step (2) of \GridSplit, we need to sort points from $\Z^d$ by
  lexicographic order. The range of the coordinates of the considered
  points is polynomial, since $G$ is connected (in fact, the range is
  linear). So we can use \emph{radix sort} to find a lexicographic
  ordering in linear time.
\end{proof}

\section{Conclusion}

We showed that any graph with edge costs and vertex weights can be
partitioned into a given number of almost equally-weighted parts in
such a way that the maximum boundary cost is small, provided that the
graph has small splittability.

Using an observation from \cite{spielman:minmax}, namely that the
boundary cost function can approximately be modeled as a (dynamic)
weight-function on the vertices, we reduced the \emph{min-max boundary
  decomposition} problem to a \emph{multi-balanced} partitioning
problem.  For the case of arbitrary edge costs, it was necessary to
balance the partition also with respect to the \emph{splitting cost
  measure}.

Finally, we developed an algorithm based on a
``shrink-and-conquer''-approach for improving the weight-balancedness
of a partition while maintaining the balanced with respect to a number
of other measures, including the boundary cost function.

We remark that, using our general framework, one can devise a
multi-balanced version of Theorem~\ref{maintheorem}: Every graph $G$
with edge costs $c$, measures $\Psi$ and $\Phi\ho 1$ through $\Phi\ho
{r}$ can be partitioned into $k$ parts such that 1.)~the $\Psi$-weight
of each part differs from the average by at most $(1-1/k)\inf \Psi$,
2.)~for each measure $\Phi\ho j$, the maximum $\Phi\ho j$-weight of
the partition is at most proportional to the average $\Phi\ho
j$-weight, and 3.) the maximum boundary cost of the partition is at
most proportional to $\sigma_p\cdot (\pnorm{c}/k^{1/p}+\Delta_c)$.

A possible direction of further work suggested by this thesis is to
investigate the question whether more general graphs have separator
theorems when one allows arbitrary edge costs.  Only planar graphs and
grid graphs are known to have separator theorems in this case.  And
our separator theorem for grid graphs still leaves space for
improvement.

\subsubsection*{Acknowledgment}
I would like to thank my advisor Peter Sanders for suggesting the
problem, and for many valuable discussions about this work.

%


\appendix

\section{Further proofs}

\subsection{Shrinking cuts}\label{app1}

In the following we shall show the lemmata needed for deriving
Corollaries~\ref{cor1}-\ref{cor3}.

Let the function $\pi\col V\ra \Rnn$ be the $p$-splitting cost measure
(cf. Definition~\ref{def:splitmeasure}) of the graph $G=(V,E)$ with
measures $\Psi$, and $\Phi\ho 1$ through $\Phi\ho r$.
\begin{lemma}\label{iterpart}
  For every $U\subseteq V$ and $\gamma\in [0,1]$ with
  $\inf{\Psi}/\Psi(U)\le \gamma/3r$, there exists a partition
  $\{X_1,\ldots,X_\ell\}$ of $U$ with $r/\gamma \le\ell\le 3r/\gamma$
  and $\bnd_UX_i=O(r/\gamma\cdot\pi^{1/p}(U))$ such that
  \[
  \frac{\gamma}{3r}\le \frac{\Psi(X_i)}{\Psi(U)}\le \frac{\gamma}{r} \
  \text{ for $i\le\ell$. }
  \]
\end{lemma}
\begin{proof}
  We just need to apply the following procedure with
  $\psi^*:=\gamma/3r\cdot \Psi(U)$.
  \begin{procedure}{\textsc{IterativePartition}}{vertex set
      $U\subseteq V$, $\psi^*\in \Rnn$}
    \irem{Precondition: $\inf\Psi\le\psi^*$}
  \item Start with $X\la U$ and $i\la 1$
  \item Until $\Psi(X)\le 3\psi^*$ repeat: \rem{Invariant: $\Psi(X)\ge
      \psi^*$} \enum{
    \item Let $X_i$ be a splitting set in $G[X]$ with $\partial_X
      X_i\le \pi^{1/p}(X)$ and $\psi^*\le \Psi(X_i)\le
      \psi^*+\inf\Psi$
    \item Update $X\la X\setminus X_i$ and increment $i\la i+1$ }
  \item Set $\ell:= i$ and $X_\ell:=X$
  \item Return the partition $\{X_1,\ldots,X_\ell\}$ of $U$
  \end{procedure}
  All parts $X_i$, especially $X_\ell$, fulfill the condition
  $\psi^*\le \Psi(X_i)\le 3\psi^*$. Hence, it holds $\ell\le
  \Psi(U)/\psi^*=3r/\gamma$ and $\ell\ge \Psi(U)/3\psi^*=r/\gamma$.
  The total cost of edges that are cut by the algorithm does not
  exceed $\ell\cdot \pi^{1/p}(U)$. It follows $\bnd_U
  X_i=O(r/\gamma\cdot \pi^{1/p}(U))$.
\end{proof}

\begin{lemma}\label{lightpart}
  For $U$ and $\gamma$ as in Lemma~\ref{iterpart}, there exists a
  subset $X$ of $U$ with $\bnd_UX=O(r/\gamma\cdot \pi^{1/p}(U))$ such
  that
  \[
  \frac{\gamma}{3r} \le \frac{\Psi(X)}{\Psi(U)}\le
  \frac{\gamma}{r},\quad \frac{\Phi\ho j(X)}{\Phi\ho j(U)}\le
  \gamma\text{ for all }j\in[r].
  \]
\end{lemma}
\begin{proof}
  By Lemma~\ref{iterpart} there exists $\ell\ge r/\gamma$ disjoint
  subsets $X_i$ of $U$ with $\Psi$-weight as required.  By the
  pigeonhole-principle, one of those parts $X_i$ has to fulfill
  $\Phi\ho j(X_i)\le \gamma\cdot \Phi\ho j(U)$ for all $j\in [r]$.
\end{proof}

The lemma below is dual to Lemma~\ref{lightpart}.

\begin{lemma}\label{heavypart}
  For $U$ and $\gamma$ as in Lemma~\ref{iterpart}, there exists a
  subset $X$ of $U$ with $\bnd_UX=O(r^2/\gamma\cdot \pi^{1/p}(U))$
  such that
  \[
  \gamma \le \frac{\Psi(X)}{\Psi(U)}\le
  \gamma+\frac{\inf\Psi}{\Psi(U)},\quad \frac{\Phi\ho j(X)}{\Phi\ho
    j(U)}\ge \frac{\gamma}{3r}\ \text{ for all }j\in[r].
  \]
\end{lemma}
\begin{proof}
  By Lemma~\ref{iterpart} there exists a partition of $U$ into parts
  $X_1,\ldots,X_\ell$ with $\ell\le 3r/\gamma$ and $\Psi(X_i)\le
  \gamma/r\cdot \Psi(U)$.  Without loss of generality we may assume
  that for all $j\in[r]$, there is a part $X_i$ with index $i\le r$
  and maximum $\Phi\ho j$-weight among all parts.  Formally, \[
  \max_{1\le i\le \min\{r,\ell\}}\!\!\!\Phi\ho j(X_i)=\max_{1\le i\le
    \ell}\Phi\ho j(X_i)\ge \Phi\ho j(U)/\ell \ \text{ for all }j\in
  [r].
  \]
  It follows for the union $\bar X:=X_1\cup \ldots X_r$ of the first
  $r$ parts that $\Phi\ho j(\bar X)/\Phi\ho j(U)\ge \gamma/3r$ for all
  $j\in[r]$.  The $\Psi$-weight of $\bar X$ cannot exceed $r\cdot
  \gamma/r\cdot \Psi(U)=\gamma\Psi(U)$. Also the cost of the boundary
  edge of $\bar X$ within $G[U]$ satisfies $\bnd_U(X)=O(r\cdot
  r/\gamma\cdot \pi^{1/p}(U))$.

  To fulfill the constraint $\Psi(X)\ge \gamma\cdot \Psi(U)$ we need
  to find a subset $S$ of $U\backslash \bar X$ such that $\Psi(\bar
  X\cup S)\ge \gamma\cdot\Psi(U)$.  So let $S\subseteq U\backslash
  \bar X$ be a splitting set in $G[U\backslash \bar X]$ with
  $\bnd_{U\backslash \bar X}(S)\le \pi^{1/p}(U)$ and
  $\gamma\cdot\Psi(U)\le \Psi(S)+\Psi(\bar X)\le
  \gamma\cdot\Psi(U)+\inf \Psi$. Then $X:=\bar X\cup S$ is a subset of
  $U$ that fulfills all requirements of the lemma.
\end{proof}

From Lemma~\ref{lightpart} and Lemma~\ref{heavypart}, we draw the
three corollaries below.  Let $\epsilon>0$ be sufficiently small and
$M:=1/\epsilon^5$, where the precise meaning of ``sufficiently''
depends only on $r$. So in Section~\ref{sec:shrink} we can assume that
$\epsilon$ and $M$ are absolute constants.  Moreover, let $\Psi^*$ be
a real number between $0$ and $\avg \Psi$ such that $\inf \Psi\le
\epsilon^5\Psi^*$. This condition corresponds to the condition on
$\inf \Psi$ in the definition of shrinking procedures
(cf. Definition~\ref{def:shrink}).

In order to achieve a geometric decrease of the boundary costs, we
choose the measure $\Phi\ho r\col V\ra \Rnn$ such that $\Phi\ho
r(v):=c(\delta(v)\cap \delta(U))$, where $U\subseteq V$ is as in the
corollaries below.  Then corollaries~\ref{cor1}-\ref{cor3} are
instantiations of the corollaries below for $r=3$.

\begin{corollary}
  For every $U\subseteq V$ with $M/2\le \Psi(U)/\Psi^*\le M$, there
  exists a subset $X$ of $U$ with $\bnd_UX=O_M(\pi^{1/p}(U))$ and
  $\epsilon \le \Psi(X)/\Psi^*\le 3\epsilon$ such that
  \begin{align*}
    \Phi\ho j(X)&\le (6r\epsilon/M)\cdot \Phi\ho j(U),\\
    \bnd X&\le (6r\epsilon/M)\cdot \bnd U+O_M(\pi^{1/p}(U)).
  \end{align*}
\end{corollary}

\begin{corollary}
  For every $U\subseteq V$ with $1/2\le \Psi(U)/\Psi^*\le M$, there
  exists a subset $X$ of $U$ with $\bnd_UX=O_M(\pi^{1/p}(U))$ and
  $\epsilon \le \Psi(X)/\Psi^*\le 3\epsilon$ such that
  \begin{align*}
    \Phi\ho j(X)&\le 6r\epsilon\cdot \Phi\ho j(U),\\
    \bnd X&\le 6r\epsilon\cdot \bnd U+O_M(\pi^{1/p}(U)).
  \end{align*}
\end{corollary}

\begin{corollary}
  For every $U\subseteq V$ with $\epsilon\le \Psi(U)/\Psi^*\le M$,
  there exists a subset $X$ of $U$ with $\bnd_UX=O_M(\pi^{1/p}(U))$
  and $\epsilon \le \Psi(X)/\Psi^*\le \epsilon+\inf\Psi/\Psi^*$such
  that
  \begin{align*}
    \Phi\ho j(U\backslash X)&\le (1-\epsilon/3r\cdot
    \frac{\Psi^*}{\Psi(U)})\cdot \Phi\ho j(U),
    \\
    \bnd (U\backslash X)&\le (1-\epsilon/3r\cdot
    \frac{\Psi^*}{\Psi(U)})\cdot \bnd U+O_M(\pi^{1/p}(U)).
  \end{align*}
\end{corollary}
\noindent We remark that one can obtain such sets $X$ as in the
corollaries above in time $O_M(t(|G[U]|))$.

\subsection{Two bin-packing procedures}\label{app2}
\begin{proof}[Proof of Lemma~\ref{conquerlem}]
  We are given two colorings $\chi_0$ and $\hat\chi_1$ of respective
  disjoint vertex sets $W_0$ and $W_1$ with $W_0\cup W_1=W$. Let
  $w^*:=w(W)/k$ denote the average weight of a $k$-coloring of
  $W$. Similarly, let $w^*_0:=w(W_0)/k$ and $w^*_1:=w(W_1)/k$ be the
  average weight of $k$-colorings in $W_0$ and $W_1$, respectively. By
  our preconditions, we have $w\ichi_0(i)=w^*_0+O(\inf w)$ and
  $w\hichi_1(i)=w^*_1+O(\inf w)$.  So it holds for every color class
  $\ichi_0(i)\cup \ichi_1(i)$ of the direct sum
  $\chi_0\oplus\hat\chi_1$, \eq{\label{eq11}
    w\ichi_0(i)+w\hichi_1(i)=w_0^*+w_1^*+O(\inf w)=w^*+O(\inf w).  }

  In the following, we write more conveniently $w_1(i):=w\hichi(i)$
  for the weight of a color class in coloring $\hat\chi_1$. We have
  the precondition $w_1(i)\le w^*-\inf w$.

  We proceed in two phases to transform $\chi_0$ into a coloring
  $\tilde\chi_0$ with the direct sum $\tilde\chi_0\oplus\hat\chi_1$
  being almost strictly balanced, i.e., $|w\tichi_0(i)+w_1(i)-w^*|\le
  2\inf w$ for all~$i\in[k]$.  We start with $\tilde\chi_0=\chi_0$.
  In color classes of $\tilde\chi_0$, we uncolor parts~$X\subseteq
  W_0$ with $\inf w\le w(X)\le 2\inf w$, until the maximum of
  $w\tichi_0(i)+w_1(i)$ over all $i\in[k]$ is at most the average
  weight $w^*$.  Then, we re-assign the previously uncolored parts $X$
  to color classes in a greedy manner. It follows that the direct sum
  $\tilde\chi_0\oplus\hat\chi_1$ is almost strictly balanced.

  Furthermore, since every considered part $X$ has weight between
  $\inf w$ and $2\inf w$, we can infer from equation~\eqref{eq11} that
  every color class of $\tilde\chi_0$ receives or emits only a
  constant number of parts. From this observation it shall follow that
  the maximum splitting cost of $\tilde\chi_0$ is at most proportional
  to the maximum splitting cost of $\chi_0$. Similarly, the maximum
  boundary cost of $\tilde\chi_0$ is in
  $O(\inf{\bnd\ichi_0}+\inf{\pi\ichi_0}^{1/p})$.

  We now give the the full details of the proof.  The procedure below
  transforms $\chi_0$ into coloring $\tilde\chi_0$.
  \begin{procedure}{ \textsc{BinPack1}}{coloring $\chi_0\col W_0\ra
      [k]$, weight function $w_1\col [k]\ra \Rnn$}
    \irem{Precondition: $w_1(i)\le w^*-\inf w$ for all $i\in [k]$}
  \item Start with $\tilde\chi_0\la \chi_0$, and $\Buffer\la
    \emptyset$.
  \item As long as there exists a color class $U=\tichi_0(i)$ %
    with $w(U)+w_1(i)>w^*$, \subitem compute a splitting set
    $X\subseteq U$ \subitem with $\inf w \le w(X)\le 2\inf w$ and
    $\bnd_U X\le \pi^{1/p}(U)$, \subitem uncolor all vertices in $X$,
    \subitem and update $\Buffer\la \Buffer\cup \{X\}$.
  \item As long as there is a color class $U=\tichi_0(i)$ with
    $w(U)+w_1(i)<w^*-2\inf w$ \subitem choose a part $X\in \Buffer$,
    \subitem paint all vertices in $X$ with color $i$, \subitem and
    update $\Buffer\la \Buffer\setminus \{X\}$
  \item For all remaining $X\in \Buffer$, \subitem choose a color
    $i\in[k]$ with $w\tichi_0(i)+w_1(i)\le w^*$, \subitem and paint
    all vertices in $X$ with color $i$.
  \item Return coloring $\tilde\chi_0$
  \end{procedure}
  First, we need to show that we really can perform the above steps as
  described.  In particular, each set $U$ selected in step $(2.)$ must
  have weight at least $\inf w$ so that a subset $X$ of $U$ with
  $w(X)\ge \inf w$ can be found.  Also, $\Buffer$ needs to be
  non-empty whenever there exists a color class with
  $w\tichi_0(i)+w_1(i)<w^*-2\inf w$ in step $(3.)$.  We also would
  have to show that there exists a color $i\in [k]$ with
  $w\tichi_0(i)+w_1(i)\le w^*$ in step $(4.)$. But this fact is
  obvious since $w^*$ is the average of $w\tichi_0(j)+w_1(j)$ over all
  $j\in[k]$.

  The invariants (I) and (II) below establish the soundness of steps
  $(2.)$ and $(3.)$ in procedure \textsc{BinPack1}.

\begin{claim}
  The following are invariants of the algorithm above.  \enum{
  \item[(I)] Every color class $U=\tichi_0(i)$ with $w(U)+w_1(i)>w^*$
    has weight $\ge w^*$
  \item[(II)] In step $(3.)$ we have $w\tichi_0(i)+w_1(i)\le w^*$ for
    all colors $i\in[k]$.  }
\end{claim}
\begin{pf}
  The precondition $w_1(i)\le w^*-\inf w$ yields $w(U)\ge \inf w$ for
  all vertex sets $U\subseteq W_0$ with $w(U)+w_1(i)\le w^*$. So
  invariant (I) holds.  Clearly, invariant (II) is valid directly
  after the completion of step $(2)$.  An iteration in step $(3)$
  maintains the invariant since $w(X)\le 2\inf w$ and therefore
  $w(U\cup X)+w_1(i)\le w^*$ for all $U\subseteq W_0$ and $i\in [k]$
  with $w(U)+w_1(i)<w^*-2\inf w$.  So invariant (II) holds, too.
  \smallqed
\end{pf}

From the following two claims, which are implied by the precondition
$w\ichi_0(i)+w_1(i)=w^*+O(\inf w)$ and the $w(X)\ge \inf w$ for all
considered parts $X$, we infer that the maximum splitting cost and the
maximum boundary cost of $\tilde\chi_0$ are as required by the lemma,
i.e., $\inf {\pi\tichi_0}=O(\inf{\pi\ichi_0}$ and
$\inf{\bnd\tichi_0}=O(\inf{\bnd\ichi_0}+\inf{\pi\ichi_0}^{1/p})$.

\begin{claim}\label{constsplit}
  The class of each color $i\in[k]$ is changed at most a constant
  number of times in steps $(2)$-$(3)$ of procedure
  $\textsc{BinPack1}$.
\end{claim}

\begin{claim}
  For each considered part $X$, we have $\pi(X)\le \inf{\pi\ichi_0}$
  and $\bnd(X)\le \inf{\bnd\ichi_0}+O(\inf{\pi\ichi_0}^{1/p})$.
\end{claim}

The claims above also imply that the procedure \textsc{BinPack1} can
be implemented to run in time at most proportional to $t(|G[W_0]|)$.
The total time for computing splitting sets is $O(t(|G[W_0]|))$ by
Claim~\ref{constsplit}. Using an appropriate data structure, e.g., a
stack, it takes constant time to select a color that satisfies a
certain condition (like $w\tichi_0(i)+w_1(i)>w^*$).

Now its easy to see that the direct sum $\tilde\chi_0\oplus\hat\chi_1$
is almost strictly balanced.  By invariant (II), we have
$w\tichi_0(i)+w\hichi_1(i)\le w*$ for all colors $i\in[k]$. When step
(3) is completed, the minimum of $w\tichi_0(i)+w\hichi_1(i)$ is at
least $w^*-2\inf w$. Since $w(X)\le 2\inf w$ for all considered parts
$X$, the maximum of $w\tichi_0(i)+w\hichi_1(i)$ cannot become larger
than $w^*+2\inf w$.
\end{proof}

\begin{proof}[Proof of Proposition~\ref{greedyprop}]

  We need the following simple claim.

\begin{claim}
  For any vertex set $W\subseteq V$ with weight at least $\inf w/2$,
  there exists $X\subseteq W$ with $\bnd_WX\le \pi^{1/p}(W)+\Delta_c$
  and $\inf w/2\le w(X)\le \inf w$.
\end{claim}
\begin{pf}
  If there exists a vertex $x\in W$ with weight at least $\inf w/2$,
  then we can choose $X:=\{x\}$.  Otherwise, we have $\inf{w\res W}\le
  \inf w/2$ and so we can compute a splitting set $X\subseteq W$ with
  $\bnd_WX\le \pi^{1/p}(W)$ and $\inf {w\res W}\le w(X)\le 2\inf{w\res
    W}$. Then part $X$ has weight between $\inf w/2$ and $\inf w$.
  \smallqed
\end{pf}

The procedure below shall compute a strictly balanced coloring
$\hat\chi$ from an almost strictly balanced coloring $\chi$ such that
$\inf{\bnd\hichi}=O(\inf{\bnd\ichi}+\inf{\pi\ichi}^{1/p}+\Delta_c)$.
Let $w^*:=\onorm w/k$ be the average weight of a $k$-coloring in $V$.
We assume $w^*\ge \inf w/2$. The somehow degenerate case $w^*<\inf
w/2$ can be handled similarly.
\begin{procedure}{\textsc{BinPack2}}{almost strictly balanced coloring
    $\chi\col V\ra [k]$}
  \irem{Precondition: $w^*\ge \inf w/2$}
\item Start with $\hat\chi\la \chi$, and $\Buffer\la \emptyset$.
\item As long as there exists a color class $U=\hichi(i)$ %
  with $w(U)>w^*$, \subitem compute a splitting set $X\subseteq U$ as
  in Claim~1 \subitem uncolor all vertices in $X$, \subitem and update
  $\Buffer\la \Buffer\cup \{X\}$.
\item As long as there is a color class $U=\hichi(i)$ with
  $w(U)<w^*-(1-1/k)\inf w$ \subitem choose a part $X\in\Buffer$,
  \subitem paint all vertices in $X$ with color $i$, \subitem and
  update $\Buffer\la \Buffer\setminus \{X\}$
\item For all remaining $X\in \Buffer$, \subitem choose a color
  $i\in[k]$ with $w\hichi(i)\le w^*-w(X)/k$, \subitem and paint all
  vertices in $X$ with color $i$.
\item Return strictly balanced coloring $\hat\chi$
\end{procedure}
Similar to Lemma~\ref{conquerlem}, in step $(3)$ the invariant
$w\hichi(j)\le w^*+\inf w/k$ holds for all colors $j\in[k]$. Hence if
there exists a color $i\in[k]$ in step $(3)$ with
$w\hichi(i)<w^*-(1-1/k)\inf w$, then the total weight of currently
colored vertices is less than $(k-1)(w^*+\inf w^*/k)+w^*-(1-1/k)\inf
w=\onorm w$.  So there are vertices uncolored and $\Buffer$ must be
non-empty.

In step $(4)$ we can choose a color $i\in[k]$ with color class of
weight at most $w^*-w(X)/k$ since the vertices in $X$ are uncolored
and so $w^*-w(X)/k$ is at least the average weight of the current
coloring $\hat\chi$.

Since $\chi$ is almost strictly balanced and each considered part $X$
has weight at least $\inf w/2$, the class of a color changes at most
constant number of times.  So for each considered part $X$, we have
$\bnd X\le \inf {\bnd\ichi}+O(\inf{\pi\ichi}^{1/p}+\Delta_c)$ by
Claim~1.  And therefore the returned coloring $\hat\chi$ satisfies
$\inf{\bnd
  \hichi}\le\inf{\bnd\ichi}+O(\inf{\pi\ichi}^{1/p}+\Delta_c)$.

The procedure \textsc{BinPack2} can be implemented to run in time
$O(t(|G|)+k\log k)$. The colors in step $(4)$ can be selected using a
heap data structure, since we can choose in every iteration the color
with class of minimum weight. This operation takes time $O(\log k)$,
since $O(k)$ parts are generated in step $(2)$.  The argument for the
running time of the remaining steps is analogous to the proof of
Lemma~\ref{conquerlem}.
\end{proof}

\subsection{Balanced Separators and Tight Examples}\label{lowersec}

In this section we elaborate on how splitting sets are related to the
more common notion of balanced separators. Specifically, we show that
both notions are equivalent for bounded-degree graphs. Based on these
results we show lower bounds for the min-max boundary decomposition
cost that are essentially proportional to the upper bounds given by
Theorem~\ref{maintheorem}.

\begin{definition}[Balanced Separation]
  A \emph{separation} of a graph $G=(V,E)$ is a pair $(A,B)$ of vertex
  sets with $A\cup B=V$ such that no edge of $G$ joins $A\setminus B$
  and $B\setminus A$.

  A separation is \emph{balanced} with respect to weights $w\colon
  V\ra \Rnn$ if the weight of both $A\setminus B$ and $B\setminus A$
  is at most two third of the total weight, i.e., $ \max\{
  w(A\setminus B),w(B\setminus A)\}\le 2/3 \cdot \onorm{w}.  $ The
  \emph{cost} of a separation $(A,B)$ with respect to a cost function
  $\tau\colon V\ra \Rnn$ is given by $\tau(A\cap B)$.
  A vertex set $S\subseteq V$ is called \emph{balanced separator} if
  $S=A\cap B$ for a balanced separation $(A,B)$.
\end{definition}
Similar to the splittability $\sigma_p$ of a graph we can define its
``separability''.
\begin{definition}[Separability, Separator Theorem]\label{def:sep}

  The \emph{$p$-separability} of $G$ with vertex costs $\tau$ is the
  minimum cost of a balanced separation in a subgraph of $G$ relative
  to the $p$-norm of the subgraph's vertex costs, where the subgraph
  and its weights are worst possible, i.e.,
  \[
  \beta_p(G,\tau):=\max_{W\subseteq V} \sup_{w\colon W\ra \Rnn}
  \min_{(A,B)}\tau(A\cap B)/\pnorm{\tau\res{W}}
  \]
  where the minimum is over all $w$-balanced separations $(A,B)$ of
  $G[W]$.

  A family $\cal G$ of pairs $(G,\tau)$ has a \emph{$p$-separator
    theorem} if there exists a constant $C_{\mathcal G}$ such that the
  $p$-separability of $G$ with vertex costs $\tau$ is at most
  $C_{\mathcal G}$ for all pairs $(G,\tau)$ in $\mathcal G$, i.e., if
  ${\beta_p}\res{\cal G}=O_{\cal G}(1)$.
\end{definition}

The remark below gives an overview of known and recent results about
the separability and splittability of various graph classes.

\begin{remark}[cf. \cite{simon97how}]\label{resultlist}
  For unit vertex costs,
  \begin{itemize}
  \item \emph{planar graphs} \cite{tarjan:sep} have $\beta_2=O(1)$,
  \item \emph{graphs with genus $g$ } \cite{gilbert:sep} have
    $\beta_2=O(\sqrt g)$,
  \item \emph{graphs excluding a clique of size $h$ as minor}
    \cite{alon:sep} have $\beta_2=O(h^{3/2})$,
  \item \emph{well-shaped meshes} in a $d$-dimensional space
    \cite{spielman:mesh} have $\beta_{d/(d-1)}=O_d(1)$,
  \item $d$-dimensional \emph{$k$-nearest neighbor graphs}
    \cite{miller:sep} have $\beta_{d/(d-1)}=O_d(k^{1/d})$.
  \end{itemize}
  For arbitrary costs,\begin{itemize}
  \item \emph{planar graphs} \cite{djidjev:cost} have $\beta_2=O(1)$,
  \item \emph{$d$-dimensional grid graphs} have $\sigma_{d/(d-1)}= O(
    d\cdot \log ^{1/d}\phi)$, where $\phi$ is the fluctuation of the
    edge costs, i.e., the ratio of the maximum cost to the minimum
    (positive) cost (cf. Section~\ref{sec:splitt-grid-graphs}).
  \end{itemize}
\end{remark}
The following lemma relates the notions of splittability and
separability.  Since we defined splittability in terms of edge costs
and separability in terms of vertex costs, we need to translate
between edge costs and vertex costs.
Let $G=(V,E)$ be a graph with edge costs $c\col E\ra \Rnn$.  A natural
choice of vertex costs $\tau\col V\ra \Rnn$ corresponding to $c$ is
given by $\tau(v):=c(\delta(v))$ for each vertex $v\in V$.  Then for
every separation $(A,B)$, the boundary cost $c(\delta(U))$ of any
vertex set $U$ with $A\backslash B\subseteq U\subseteq A$ is no more
than $\tau(A\cap B)$.
On the other hand, we want to be able to construct from vertex sets
$U'\subseteq V$ separations $(A',B')$ with $U'\subseteq A'$ and cost
$\tau(A',B')$ proportional to $c(\delta(U'))$.  For this, we require
that the \emph{local fluctuation} $\lfluc(c):=\max_{u\in e\in
  E}\tau(u)/c(e)$ is bounded.  When we choose $B':=V\backslash U$ and
$A'$ to be the set of vertices reachable from $U'$ by at most one
edge, then the separation $(A',B')$ has cost at most $\tau(A'\cap
B')\le2\lfluc(c)\cdot c(\delta(U'))$, since every vertex $A'\cap B'$
is an endpoint of an edge in $\delta(U')$.

Notice that for the case of unit edge costs, the local fluctuation
$\lfluc(\bbbone)$ equals the maximum degree $\Delta$ of $G$.

In the following, we shall see that $\sigma_p(G,c)$ is proportional to
$\beta_p(G,\tau)$ when both the maximum degree $\Delta$ and the local
fluctuation $\lfluc$ are bounded.  The lemma's proof is a slight
generalization of the proof in \cite{tarjan:sep} for the fact that
cheap balanced separations imply inexpensive separations $(A,B)$ with
both $w(A\setminus B)$ and $w(B\setminus A)$ at most $\onorm w/2$.
\begin{lemma}\label{balsep}
  Let $G=(V,E)$ be a graph with edge costs $c\colon E\ra \Rnn$.
  Then \[ \beta_p(G,\tau)/\lfluc(c) \stackrel{1.)}{\ll_p}
  \sigma_p(G,c) \stackrel{2.)}{\ll_p} \lfluc\cdot\Delta^{1/q} \cdot
  \beta_p(G,\tau)\] where $\tau$ are vertex costs with
  $\tau(v):=c(\delta(v))$, $f\ll_p g$ is short for $f=O_p(g)$, and
  $\frac{1}{p}+\frac{1}{q}=1$.
\end{lemma}
\begin{proof}
  \noindent\emph{1.)} $\beta_p=O(\lfluc\cdot \sigma_p)$:
  Let $W$ be a subset of $V$ with weights $w\col W\ra \Rnn$.  We need
  to show that there exists a $w$-balanced separation $(A,B)$ of cost
  $\tau(A\cap B)=O(\lfluc\cdot \sigma_p\cdot \pnorm{\tau\res W})$.

  If $w(v)> \onorm w/3$ for some vertex $v\in W$ then $(\{v\},W)$ is a
  $w$-balanced separation of cost $\tau(v)\le \pnorm{\tau\res W}$.

  So we can assume $\inf w\le \onorm w/3$.  Then let $U\subseteq W$ be
  a splitting set with $\partial_W U\le \sigma_p\cdot\pnorm{c\res W}$
  and $1/3\cdot \onorm w\le w(U)\le 1/3\cdot \onorm w+\inf w$.  Our
  assumption ensures $w(U)\le 2/3\cdot \onorm w$.  Let $X\subseteq W$
  contain the endpoints of the edges in the cut $C:=\delta_{G[W]}(U)$.
  Now $(A,B):=(U\cup X,W\setminus U)$ is a balanced separation of
  $G[W]$.  The cost of $(A,B)$ satisfies \begin{align*} \tau(A\cap
    B)&\le\tau(X) \le \sum\nolimits_{\{u,v\}\in C}\tau(u)+\tau(v)
    =\sum\nolimits_{e\in C}2\lfluc\cdot c_e\\
    &=O(\lfluc \cdot c(C))=O(\lfluc\cdot \sigma_p\cdot \pnorm{c\res
      W})
  \end{align*}
  And the $p$-norm of $c\res W$ is at most proportional to the
  $p$-norm of $\tau\res W$, since \[ 2\cdot\sum_{e\in E[W]}
  c^p_e=\sum_{v\in W} \sum_{e\in \delta(v)} c^p_e \le \sum_{v\in
    W}(\tau(v))^p
  \]
  So $(A,B)$ is a balanced separation with cost at most proportional
  to $\lfluc\cdot \sigma_p\cdot \pnorm{\tau\res W}$.

  \emph{2.)} $\sigma_p=O_p(\lfluc\cdot \Delta^{1/q}\cdot \beta_p)$:
  Let $W$ be a subset of $V$ with weights $w\col W\ra \Rnn$ and
  splitting value $w^*$.  We need to show that there exists
  $w^*$-splitting set $U\subseteq W$ of cost $\bnd_W U
  =O_p(\lfluc\cdot \Delta^{1/q} \cdot \beta_p\cdot \pnorm{c\res{W}})$.

  Similar to the $p$-splitting cost measure
  (cf. Definition~\ref{def:splitmeasure}), we define a weight function
  $\pi\col V\ra \Rnn$ with $\pi(v):=(\beta_p\cdot
  c(\delta(v))^p=(\beta_p\cdot \tau(v))^p$.  Then it holds
  $\pi(W')=\beta_p\pnorm{\tau\res{ W'}}$ for arbitrary vertex set
  $W'\subseteq V$.  Thus, there exists balanced separators of cost
  $\pi^{1/p}(W'):=(\pi(W'))^{1/p}$ in $G[W']$, and so we can call
  $\pi^{1/p}(W')$ the \emph{separating cost} of $G[W']$.

  The following procedure computes a separation $(A_0,B_0)$ of $G[W]$
  such that $w(A_0\backslash B_0)\le w^*-\inf w/2\le w(A_0)$.  The
  idea is to divide the vertices of $G[W]$ using a $\pi$-balanced
  separation $(A,B)$ and then to proceed recursively on one of the
  graphs $G[A\setminus B]$ and $G[B\setminus A]$.
  \begin{procedure}{Split}{vertex set $W\subseteq V$, $w^*\in \Rnn$} %
    \irem{Precondition: $0\le w^*\le \onorm{w\res{W}}$}
  \item \emph{Trivial case:} if $\pi(W)=0$, then return separation
    $(W,W)$
  \item Let $(A,B)$ be $\pi$-balanced separation of $G[W]$ with cost
    $\tau(A\cap B)\le \beta_p\pnorm{\tau\res{W}}=\pi^{1/p}(W)$
  \item If $w^*-\inf w/2<w(A\setminus B)$ \subitem then let
    $(A',B')=\textsc{Split}(A\setminus B,w^*)$ be separation of
    $G[A\setminus B]$\subitem and return $(A_0,B_0):=(A'\cup (A\cap
    B), B'\cup B)$,
  \item else if $w(A\setminus B)\le w^*-\inf w/2\le w(A)$ \subitem
    then return $(A_0,B_0):=(A, B)$,
  \item else if $w(A)<w^*-\inf w/2$ \subitem then let
    $(A',B')=\textsc{Split}(B\setminus A,w^*-w(A))$ be separation of
    $G[B\setminus A]$ \subitem and return $(A_0,B_0):=(A\cup A',B'\cup
    (A\cap B))$.
  \end{procedure}
  Since both $\pi(A\setminus B)$ and $\pi(B\setminus A)$ are at most
  $\frac{2}{3}\cdot \pi(W)$, it follows by induction on the size of
  the considered graph that $\tau(A_0\cap B_0)=\tau(A\cap
  B)+\tau(A'\cap B') \le
  \pi^{1/p}(W)\cdot\sum_{i=0}^\infty(\frac{2}{3})^{i/p}=O_p(\pi^{1/p}(W))$.

Without loss of generality we may assume that $G[W]$ is connected. 
(If $G[W]$ was not connected, we would need to apply $\textsc{Split}$
only to one of the connected components of $G[W]$.)
Hence it holds $\tau(v)=c(\delta(v))\le \lfluc\cdot
c(\delta(v)\cap E(W))=c(\delta_{G[W]}(v))$. Then we have by H\"older's inequality 
\[
\tau(v)=\lfluc \cdot \sum_{e\in\delta_{G[W]}(v)} c_e \le \lfluc \cdot
|\delta_{G[W]}(v)|^{1/q}\cdot (\sum_{e\in\delta_{G[W]}(v)}c^p_e)^{1/p}
\le\lfluc \cdot \Delta^{1/q} \cdot
(\sum_{e\in\delta_{G[W]}(v)}c^p_e)^{1/p}.
\] So it holds $\pi(W)\le
\lfluc \cdot \Delta^{1/q} \cdot \sum_{v\in
  W}\sum_{e\in\delta_{G[W]}(v)} c^p_e$ and also $\pi^{1/p}(W)=O(\lfluc
\cdot\Delta^{1/q}\cdot \beta_p\cdot \pnorm {c\res W})$.  Thus we get
$\tau(A_0\cap B_0)=O_p(\lfluc\cdot\Delta^{1/q}\cdot \beta_p\cdot
\pnorm{c\res W})$.

Given a separation $(A_0,B_0)$ computed by \textsc{Split}($W$, $w^*$),
we find a $w^*$-splitting set $U$ of $G[W]$ as follows.  Let
$\{v_1,\ldots,v_h\}=A_0\cap B_0$ be an enumeration of the separator
$A_0\cap B_0$, and let $i\in[h+1]$ be the largest index with
$w(A\setminus B)+w(v_1)+\ldots+w(v_{i-1})\le w^*-\inf w/2$.  Then
$U:=A_0\setminus B_0\cup \{v_1,\ldots,v_{i-1}\}$ is $w^*$-splitting.

Since $A_0\setminus B_0\subseteq U\subseteq A_0$, the boundary cost of
$U$ cannot exceed $\tau(A_0\cap B_0)$ and it holds $\bnd
U=O_p(\pi^{1/q}(W))=O_p(\lfluc \cdot \Delta^{1/q}\cdot \beta_p\cdot
\pnorm{c\res W})$ as required.
\end{proof}

We remark that the running time of procedure \textsc{Split} might be
quite long.  The reason is that the size of one of the graphs
$G[A\setminus B]$ and $G[B\setminus A]$ could be almost as large as
$|G[W]|$. However this issue can be resolved, by using for every
second recursive call of the procedure (alternately), separations that
are $deg_W$-balanced instead of $\pi$-balanced, where $deg_W\col V\ra
\Rnn$ assigns the degree in $G[W]$ to a vertex. Then both graphs
$G[A\setminus B]$ and $G[B\setminus A]$ have size at most
$\frac{2}{3}|G[W]|$.  With this modification, \textsc{Split}($W$,
$w^*$) runs in time $O(t(|G[W]|))$, provided that one can find
balanced separations of graphs $G[W']$ in time $t(|G[W']|)$ and $t\col
\N\ra \N$ is a linear function.

Notice that the second part of the proof of Lemma~\ref{balsep} implies
the following stronger statement.  Let $b$ be the maximum of
$\min_{(A,B)} \tau(A\cap B)/\pnorm{\tau\res W}$ over all sets
$W\subseteq V$, where the minimum is over all $\pi$-balanced
separations of $G[W]$.  Then for arbitrary weights $w$, procedure
\textsc{Split} can find $w$-balanced separations of $G$ with cost at
most $O_p(b\cdot \pnorm{\tau})$.  Hence, we can draw the corollary
below from the second part of the proof of Lemma~\ref{balsep}
(cf. \textsc{Split}). This corollary observes that cheap balanced
separators with respect to one ``universal'' measure imply cheap
balanced separators with respect to arbitrary measures.
\begin{corollary}
  Let $G=(V,E)$ be a graph with vertex costs $\tau$, and weights $\pi$
  be as in the proof of Lemma~\ref{balsep}.
  Then, \[\beta_p(G,\tau)\ll_p \max_{W\subseteq V} \min_{(A,B)}
  \tau(A\cap B))/\pnorm{\tau\res W}\] where the minimum is over all
  $\pi$-balanced separations of $G[W]$.
\end{corollary}

Similar to the situation above, the first part of the proof of
Lemma~\ref{balsep} implies the following stronger statement.  Let
$s:=\max_{W\subseteq V}\bnd^2_\infty(G[W],c\res W)/\pnorm{c\res W}\ge
\sigma_p$.  If we used strictly balanced $2$-colorings instead of
splitting sets, we could show $\beta_p=O(\lfluc\cdot s)$.  Together
with the second part, we obtain an upper bound on $\sigma_p$ in terms
of the min-max boundary decomposition cost $\partial^k_2$ for two
colors:
\begin{corollary}\label{splitcol}
  For graphs $G=(V,E)$ with edge costs $c\col E\ra \Rnn$, it holds
  \[\sigma_p(G,c)\ll_p\Delta^{1/q}\cdot\lfluc^2(c)\cdot
  \max_{W\subseteq V}\bnd^2_\infty(G[W],c\res W)/\pnorm{c\res W}
  \le\Delta^{1/q}\cdot\lfluc^2(c)\cdot \sigma_p(G,c).\]
\end{corollary}

In the remainder of this section 
we construct families of instances for which we can compute
lower bounds on the min-max boundary decomposition cost. 
With these instances, we can argue that there is no way to improve
the upper bound of Theorem~\ref{thm:2}, i.e., the bound
is optimal with respect to the chosen parameters.

The idea is as follows. Let $G=(V,E)$ be a graph with edge costs $c$
of bounded local fluctuation and weights $w$ such that each balanced
separation has cost $\Omega(b\cdot\pnorm{\tau})$.  Then consider the
graph $\tilde G=G\ho 1\dot\cup \ldots \dot\cup G\ho {\lfloor
  k/4\rfloor}$ consisting of $\floor{k/4}$ disjoint isomorphic copies
$G\ho i$ of $G$.  For a vertex $v\in V$, we write $v\ho i$ for the
copy of $v$ in $G\ho i$.  Similarly, $e\ho i$ denotes the edge in
$G\ho i$ that corresponds to an edge $e\in E$.  We extend the costs
and weights of $G$ to the graph $\tilde G$ in the obvious way: $\tilde
c(e\ho i):=c(e)$ and $\tilde w(v\ho i):=w(v)$.  Now the claim is that
every $k$-coloring $\chi$ of $\tilde G$ with $\inf{\tilde w\ichi}\le
2\avg{\tilde w}$ has average boundary cost $\Omega(b\cdot
k^{-1/p}\cdot \pnorm{\tilde c})$.

\begin{lemma}\label{lowbnd1}
  Let $k\ge 4$ be an integer and $G=(V,E)$ be a graph with edge costs
  $c\col E\ra \Rnn$ and vertex weights $w\col V\ra \Rnn$.  Suppose all
  $w$-balanced separations of $G$ have cost at least $b\cdot
  \pnorm{\tau}$ with respect to the vertex costs $\tau$ that
  correspond to $c$, i.e., $\tau(v):=c(\delta(v))$.

  Then for the graph $\tilde G=(\tilde V,\tilde E)$ that consists of
  $\floor{k/4}$ pairwise disjoint isomorphic copies of $G$, every
  $k$-coloring $\chi$ with roughly balanced weights, i.e.,
  $\inf{\tilde w\ichi}\le 2\avg{\tilde w}$, has average boundary cost
  \[
  \avg{\bnd\ichi}=\Omega(b\cdot k^{-1/p}\cdot \pnorm {\tilde
    c}/\lfluc(c)),
  \]
  where $\tilde c\col \tilde E\ra \Rnn$ and $\tilde w\col \tilde V\ra
  \Rnn$ are the extensions of $c$ and $w$ to $\tilde G$, respectively.
\end{lemma}
\begin{proof}
  We consider one of the isomorphic copies of $G$, say $G\ho i=(V\ho
  i,E\ho i)$.  Let $U_j:=\ichi(j)\cap V\ho i$ be the set of vertices
  of $G\ho i$ with color $j$ in coloring $\chi$.
 
  The coloring $\chi$ has maximum weight at most $2\avg{\tilde w}$,
  and the weight of $G\ho i$ is at least $4\avg{\tilde w}$.  Hence for
  each $U_j$, it holds $w(U_j)\le w(V\ho i)/2$.  Then we can
  (greedily) find a partition $\{R,B\}$ of the color set $[k]$ such
  that $\sum_{j\in R}w(U_j)\le 2/3\cdot w(V\ho i)$ and $\sum_{j\in
    B}w(U_j)\le 2/3\cdot w(V\ho i)$.

  Let $U^*\subseteq V\ho i$ denote the set $\bigcup_{j\in R}U_j$ and
  $X\subseteq V\ho i\setminus U^*$ be the set vertices reachable from
  $U^*$ by exactly one edge (of $\delta(U^*)$).  So $(A,B):=(U^*\cup
  X,V\ho i\setminus U^*)$ is a $w$-balanced separation of $G\ho i$.
  By our precondition, we know $\tau(A\cap B)=\tau(X)\ge b\cdot
  \pnorm{\tau}$.  As in the first part of the proof of
  Lemma~\ref{balsep}, we have $\bnd U^*=\Omega(\tau(A\cap B)/\lfluc)$
  and $\pnorm{\tau}=\Omega(\pnorm{c})$.

  Therefore, the total boundary cost of the coloring $\chi\res {V\ho
    i}$ satisfies $\onorm{\bnd\ichi\res {V\ho i}}\ge \bnd(U^*)
  =\Omega(b\cdot \pnorm{c}/\lfluc)$.  Since $i$ was arbitrary, the
  total boundary cost of $\chi$
  \[\onorm{\bnd\ichi}=\sum_{i=1}^{\floor{k/4}}\onorm{\bnd\ichi\res
    {V\ho i}}\ge \floor{k/4}\cdot \Omega(b\cdot \pnorm{c}/\lfluc)\]
  and hence the average boundary cost of $\chi$ is at least
  proportional to $b\cdot \pnorm{c}/\lfluc$.

  Since $\pnorm{\tilde c}^p=(\sum_i\sum_{e\in E}c^p_e)
  =\floor{\frac{k}{4}}\pnorm{c}^p$, it holds $ \pnorm{c}=\Omega(
  \pnorm{\tilde c}/k^{1/p})$ and thus we have
  $\avg{\bnd\ichi}=\Omega(b\cdot k^{-1/p} \cdot \pnorm{\tilde
    c}/\lfluc)$ as required.
\end{proof}

We get the following corollary from Lemma~\ref{lowbnd1}.
Any graph for which we know a lower bound on the minimum cost balanced
separation allows us to construct a ``similar'' graph with a lower
bound on $\partial^k_\infty$ that matches the upper bound from
Theorem~\ref{thm:2}.  We assume that the instance is well-behaved,
i.e., the graph has bounded maximum degree $\Delta$ and the edge costs
have bounded local fluctuation $\lfluc$.

\begin{corollary}\label{cor:c1}
  Let $G=(V,E)$ be a well-behaved graph with edge costs $c\col E\ra
  \Rnn$ and a $p$-separator theorem (with respect to vertex costs
  $\tau(v):=c(\delta(v))$ that correspond to the edge costs).  Suppose
  there are weights $w\col V\ra \Rnn$ such that $\inf w\le \onorm w/4$
  and all $w$-balanced separations of $G$ have cost
  $\Omega(\pnorm{\tau})$.

  Then for every positive multiple of $4$, say $k$, there exists a
  well-behaved graph $\tilde G$ with edge costs $\tilde c$ and a
  $p$-separator theorem such that
  \begin{equation}
    \label{eq:13}
    \bnd_\infty^k(\tilde G,\tilde c)=\Theta_p(\pnorm{\tilde c}/k^{1/p}+\inf{\tilde c}).
  \end{equation}

  Also there are weights $\tilde w$ of $\tilde G$ such that every
  roughly $\tilde w$-balanced coloring has average boundary cost
  $\Omega(\pnorm{\tilde c}/k^{1/p}+\inf {\tilde c})$.
\end{corollary}

\begin{proof}
  Let $\tilde G$ and $\tilde c$ be as in Lemma~\ref{lowbnd1}.  Observe
  that $\inf {\tilde c}\le \pnorm{c}=O(\pnorm{\tilde c}/k^{1/p})$.

  The condition $\inf w\le \onorm w/4$ ensures $\avg{\tilde
    w}=\floor{k/4} \frac{\onorm w}{k}=\onorm w/4\ge \inf w$ and
  therefore every strictly $\tilde w$-balanced coloring of $\tilde G$
  has maximum weight at most $\avg{\tilde w}+\inf{ \tilde w}\le 2\avg
  {\tilde w}$.

  So it follows from Lemma~\ref{lowbnd1} that the average boundary
  cost (and also the maximum boundary cost) of every roughly or
  strictly balanced coloring is $\Omega(\pnorm{\tilde c}/k^{1/p}+\inf
  {\tilde c})$.

  The instance $(\tilde G,\tilde c)$ is well-behaved and has a
  $p$-separator theorem, because it is a disjoint union of
  well-behaved instances with $p$-separator theorem.  In fact, it is
  an easy consequence of procedure $\textsc{Split}$
  (cf. Lemma~\ref{balsep}) that $(\tilde G,\tilde c)$ has a
  $p$-separator theorem.

  Then it follows from Lemma~\ref{balsep} and the well-behavior of $G$
  that the $p$-splittability of $\tilde G$ is at most a
  constant. Since $G$ has bounded maximum degree, it holds
  $\Delta_c(\tilde G)=O(\inf {\tilde c})$.  So by
  Theorem~\ref{maintheorem} we have
  \[
  \bnd_\infty^k(\tilde G,\tilde c)=O_p(\pnorm{\tilde
    c}/k^{1/p}+\inf{\tilde c}). \qedhere
  \]
\end{proof}

We remark that it is a common assumption in previous work
\cite{spielman:minmax} to require from the considered graphs a
$p$-separator theorem and bounded degree, i.e., that $\beta_p$ and
$\Delta$ are constants.  We need the additional assumption that
$\lfluc$ is bounded, since we consider arbitrary costs instead of only
unit-costs.  Recall that $\lfluc=\Delta$ for unit costs.

The arguments in the proofs of Lemma~\ref{lowbnd1} and
Corollary~\ref{cor:c1} yield Theorem~\ref{thm:2}.

\end{document}